\documentclass{aastex63}

\usepackage{enumitem}
\usepackage{multirow}
\usepackage{amsmath}
\usepackage{booktabs}
\usepackage{blindtext}
\usepackage{placeins}
\received{January 25, 2021}
\revised{April 23, 2021}
\accepted{April 26, 2021}
\submitjournal{The Planetary Science Journal}

\graphicspath{{./}{figures/}}

\begin{document}

\title{Chandrayaan-2 Dual-Frequency SAR (DFSAR):
Performance Characterization and Initial Results}

\correspondingauthor{Sriram S. Bhiravarasu}
\email{sriram.saran@sac.isro.gov.in, sriram.saran@gmail.com}

\author[0000-0003-0019-6261]{Sriram S. Bhiravarasu} 
\affiliation{Space Applications Centre, ISRO, 
Ahmedabad 380015, India}
\author {Tathagata Chakraborty}
\affiliation{Space Applications Centre, ISRO, 
Ahmedabad 380015, India}
\author {Deepak Putrevu}
\affiliation{Space Applications Centre, ISRO, 
Ahmedabad 380015, India}
\author {Dharmendra K. Pandey}
\affiliation{Space Applications Centre, ISRO, 
Ahmedabad 380015, India}
\author {Anup K. Das}
\affiliation{Space Applications Centre, ISRO, 
Ahmedabad 380015, India}
\author {V. M. Ramanujam}
\affiliation{Space Applications Centre, ISRO, 
Ahmedabad 380015, India}
\author {Raghav Mehra}
\affiliation{Space Applications Centre, ISRO, 
Ahmedabad 380015, India}
\author {Parikshit Parasher}
\affiliation{Space Applications Centre, ISRO, 
Ahmedabad 380015, India}
\author {Krishna M. Agrawal}
\affiliation{Space Applications Centre, ISRO, 
Ahmedabad 380015, India}
\author {Shubham Gupta}
\affiliation{Space Applications Centre, ISRO, 
Ahmedabad 380015, India}
\author {Gaurav S. Seth}
\affiliation{Space Applications Centre, ISRO, 
Ahmedabad 380015, India}
\author {Amit Shukla}
\affiliation{Space Applications Centre, ISRO, 
Ahmedabad 380015, India}
\author {Nikhil Y. Pandya}
\affiliation{Space Applications Centre, ISRO, 
Ahmedabad 380015, India}
\author {Sanjay Trivedi}
\affiliation{Space Applications Centre, ISRO, 
Ahmedabad 380015, India}
\author {Arundhati Misra}
\affiliation{Space Applications Centre, ISRO, 
Ahmedabad 380015, India}
\author {Rajeev Jyoti}
\affiliation{Space Applications Centre, ISRO, 
Ahmedabad 380015, India}
\author {Raj Kumar}
\altaffiliation{Currently at: National Remote Sensing Centre, ISRO, 
Hyderabad 500037, India}
\affiliation{Space Applications Centre, ISRO, 
Ahmedabad 380015, India}

\begin{abstract}

The Dual-Frequency synthetic aperture radar (DFSAR) system manifested on the Chandrayaan-2 spacecraft represents a significant step forward in radar exploration of solid solar system objects. It combines SAR at two wavelengths (L- and S-bands) and multiple resolutions with several polarimetric modes in one lightweight ($\sim20$ kg) package. The resulting data from DFSAR support calculation of the 2$\times$2 complex scattering matrix for each resolution cell, which enables lunar near surface characterization in terms of radar polarization properties at different wavelengths and incidence angles. In this paper, we report on the calibration and preliminary performance characterization of DFSAR data based on the analysis of a sample set of crater regions on the Moon. Our calibration analysis provided a means to compare on-orbit performance with pre-launch measurements and the results matched with the pre-launch expected values. Our initial results show that craters in both permanently shadowed regions (PSRs) and non-PSRs that are classified as Circular Polarization Ratio (CPR)-anomalous in previous S-band radar analyses appear anomalous at L-band also. We also observe that material evolution and physical properties at their interior and proximal ejecta are decoupled. For Byrgius C crater region, we compare our analysis of dual-frequency radar data with the predicted behaviours of theoretical scattering models. If crater age estimates are available, comparison of their radar polarization properties at multiple wavelengths similar to that of the three unnamed south polar crater regions shown in this study may provide new insights into how the rockiness of craters evolves with time.

\end{abstract}
\section{Introduction} \label{sec:1}
The physical properties of lunar surface materials provide an independent means to help unravel lunar geologic history and evaluate their exploration potential. Additionally, near surface physical properties of the regolith are of importance to address significant questions in lunar geology, such as understanding the spatial limits and relative timing of mare flow units and impact related processes that eventually influence spatial differences in surface and subsurface composition (e.g. \cite{Cahill_2014}). In this context, polarimetric radar backscatter data for the Moon provides valuable insights regarding the physical properties of the near surface with particular sensitivity to ilmenite content, and surface or buried rocks with diameter of about one-tenth the radar wavelength and larger (e.g. \cite{Carter_2009, Campbell_2010, Campbell_2012}).

The Dual-Frequency Synthetic Aperture Radar (DFSAR) aboard India’s second lunar mission Chandrayaan-2 is the first fully polarimetric synthetic aperture radar (SAR) outside Earth orbit. The architecture of this radar instrument supports multiple polarimetric modes of operation \citep{Putrevu_2016, Putrevu_2020}, and illustrates the value of dual-frequency quad polarimetry for lunar and planetary applications. Regarding applications to the lunar surface, the main novelties provided by this instrument are the following:

\textit{Frequency bands}: For the first time ever, it is possible to analyse L- band ($\sim$24 cm wavelength) polarimetric radar images of the Moon from the DFSAR instrument. Compared to previous ground-based (e.g. Arecibo, Goldstone) and orbital-based (e.g. Mini-SAR and Mini-RF) radar data of the Moon collected at 70-cm (P band), 12.6-cm (S band) and 3.8-cm (X band) wavelengths, the new L-band SAR allows for a penetration depth of $\sim$3 meters (for dry, low-loss soils) with particular sensitivity to ilmenite content and surface or buried rocks with diameter of about one-tenth the radar wavelength (i.e. $\sim$2 cm) and larger. Additionally, combined with the S-band radar mode of operation (12 cm wavelength) along with data from the Mini-SAR and Mini-RF radars on board the Chandrayaan-1 and LRO missions respectively \citep{Nozette_2010}, the DFSAR instrument can characterize the radar scattering properties of a top few meters of the lunar surface.

\textit{Full-polarization}: The DFSAR architecture introduced the fully polarized (FP) case among polarimetric imaging radars for planetary missions, in which the intrinsic data product is the 4$\times$4 scattering matrix of each resolved element in the scene. FP systems alternately transmit two orthogonal polarizations and record both received polarizations (HH, HV, VH, and VV), and allow much more information to be extracted from a scene compared with single- and dual-pol SAR data \citep{van_Zyl_1987}. For each combination, the first letter refers to the transmitted polarization, while the second one refers to the received sense with H and V denoting horizontal and vertical respectively. After applying certain symmetry relations, this can be reduced to a 3$\times$3 array such as the compressed Sinclair matrix or the compressed Stokes matrix \citep{LeeJong-Sen2009}; such reduced forms are referred to as quadrature-polarimetric SAR (quad-pol for short). Polarimetric channels acquired in quad-pol mode maintain their relative phase, so they can be combined coherently to form new channels \citep{LeeJong-Sen2009} or to compute statistical higher order parameters by target decompositions (e.g. \cite{Cloude_1996}). Note that in the polarimetric radar literature, the terms “FP” and “quad-pol” are often used synonymously, so we adopt similar notation throughout this paper. Quad-pol data produces a unique scattering matrix using any combination of transmit and receive orthogonal polarizations (e.g. linear, elliptical, hybrid), which permits analysis of the surface scattering behaviour in all possible configurations of the transmitted and received signal polarization. Such uniqueness does not apply to single, dual, and compact polarimetric SAR configurations (e.g. \cite{LeeJong-Sen2009})

\textit{Spatial resolution and look geometry}: To address a variety of scientific objectives, DFSAR collects data of the lunar surface at a wide range of slant-resolution options spanning 2–75 m per pixel. The high spatial resolution of DFSAR enables mapping of lunar craters and other geological features, especially in the polar regions, with finer details. Compared to previous orbital and ground-based radars, high spatial resolution DFSAR data also provide sufficient number of samples for the investigation of distributed scatterers that are common in lunar geologic settings. Radar backscatter will also vary depending on the incidence angle of the radar beam. Previous studies using FP SAR data of terrestrial settings indicated that surface scattering models could be developed using the relationship between surface roughness and dielectric parameters and the intensity, polarization, and angular dependence of the backscattered wave (e.g. \cite{Campbell_1993, Campbell_1996, Campbell_2002}. In addition, theoretical models of radar scattering predict that at the near-nadir regime (i.e. from normal incidence to $\sim20^{\circ}$), the reflection is dominated by locally smooth, radar-facing facets of the surface, and for incidence angles beyond $\sim30^{\circ}$ diffuse scattering from the small-scale roughness predominates \citep{Hagfors_1964, Farr_1993}. The DFSAR instrument is designed to collect data at a wide range of angles of incidence, from 9.6$^{\circ}$ to 36.9$^{\circ}$ \citep{Putrevu_2016, Putrevu_2020}). This capability enables acquisition of dual-wavelength (L- and S-bands) radar images at multiple viewing angles to address some important applications such as the estimation of wavelength-scale physical properties (e.g. surface root mean square height and slope) \citep{Campbell_1993, Campbell_1996}, and to reduce the ambiguities related to the interpretation of high Circular Polarization Ratios (CPR) exhibited by certain lunar surface features \citep{Campbell_2012, Spudis_2013, Fassett_2018, Fa_2018, Virkki_2019}.\\

The text is organized as follows. Section \ref{sec:2} describes the scientific background and objectives of DFSAR investigations. Section \ref{sec:3} is dedicated to the instrument overview and calibration of DFSAR data products. A brief description of DFSAR polarimetric parameters along with their physical interpretation is discussed in section \ref{sec:4}. In section \ref{sec:5}, we discuss about the DFSAR data used in this preliminary analysis. Results are presented in section \ref{sec:6}, followed by conclusions and future work in section \ref{sec:7}.
\section{Background of Investigation}\label{sec:2}
\textit{Polar Volatiles}: The nature and distribution of volatiles (e.g. water ice) at the permanently shadowed regions (PSRs) of the Moon has been a subject of considerable controversy, due to contrasting interpretations of the polarimetric behaviour of the radar backscatter from these regions at near-zero phase (bistatic) angles (e.g. \cite{Fa_2013, Nozette_1996, Campbell_2006, Spudis_2010, Spudis_2013, Fa_2013, Eke_2014, Fa_2018, Fassett_2018}). The presence of ice in the regolith causes a distinct, but not unique, CPR signature caused due to forward scattering of radar signals by cracks and voids in the ice, which has an intrinsically low microwave loss. Due to the low-loss properties of ice, CPR values above 1 have been observed for the Galilean satellites \citep{Campbell_1978, Ostro_1992, Black_2001}, and the Greenland ice sheet \citep{Rignot_1995}. Within some polar craters of Mercury, features exhibiting a very strong radar backscatter and enhanced CPR values along with a high degree of correlation between radar-bright features and regions of permanent shadow have been attributed to several radar wavelengths-thick slabs of water ice \citep{Harmon_1994, Harmon_2001, Harmon_2011, Chabot_2012, Chabot_2013, Chabot_2018}. One of the most plausible mechanisms that could produce these high backscatter and CPR values is the coherent backscatter model \citep{Hapke_1990}, which requires scattering centres (cracks or inhomogeneities) embedded in a low-loss matrix such as ice \citep{Hapke_1991, Mishchenko_1992}. Unlike the Galilean satellites and Mercury poles, radar observations of lunar PSRs did not show radar-bright regions that indicates the presence of large expanses of water ice but indicated that the possible ice could be present as few wt\% in the uppermost meter of regolith or mixed as patches within the polar regolith as “dirty ice” \citep{Nozette_1996, Thompson_2011, Thomson_2012}. If, however, the lunar ice is present in the regolith in the form ice‐filling pores, then radar scattering differences might be too small to detect \citep{Thompson_2011, Virkki_2019}. Another line of evidence from the interpretation of Mini-SAR and Mini-RF radar data suggested that both poles contain abundant water in the form of relatively “clean” ice, within the upper couple of meters of the lunar surface \citep{Spudis_2010, Spudis_2013}. Other work has called this interpretation into question and have interpreted PSRs with enhanced CPR signatures as indicative of effects due to roughness, not water ice \citep{Simpson_1999, Campbell_2006, Fa_2013, Eke_2014, Fa_2018}.

\textit{Impact ejecta and melts}: Polarimetric radar is a powerful tool for studying lunar impact craters, exposing their “rough” ejecta deposits and associated impact melts \citep{Campbell_2012, Neish_2014}. New multi-wavelength radar observations would enable constraining block sizes and to distinguish surface from buried rocks in proximal and distant ejecta deposits \citep{Ghent_2016, Campbell_2002}. The impact melt deposits have very high CPRs compared to other features on the Moon, suggesting that their surfaces are some of the roughest material on the Moon at the centimetre to decimetre scale, even though they appear smooth at the meter scale (e.g. \cite{Carter_2012, Neish_2014, Neish_2017}).

\textit{Radar-dark haloes}: A certain class of impact craters on the Moon have radar-dark haloes around them, comprised of fine-grained, rock-poor ejecta distal to the blocky proximal ejecta \citep{Ghent_2005}. The composition of this halo material is suggested to be a mixture of crater ejecta and pre-existing regolith (e.g. \cite{Ghent_2016}), and radar data at multiple wavelengths could provide significant new insight into the behaviour of ejecta and their effects on the surrounding terrain.
\subsection{DFSAR objectives}\label{sec:2.1}
After end of commissioning phase in September 2019, science mission phase of Chandrayaan-2 has started which will continue till any major contingencies hamper the spacecraft operation and/or science data reception. The DFSAR instrument is designed to address key science questions related to the aforementioned target properties during the Chandrayaan-2 science mission phase, by probing the lunar regolith to understand surface physical and dielectric properties, structure, and exploring the PSRs at the lunar poles. Specifically, DFSAR data will be utilized in characterizing the physical properties of lunar mare, impact craters and their associated ejecta and melt deposits, volcanic deposits including pyroclastic mantling material and domes, buried layering, and volatiles at lunar poles.

The DFSAR architecture is new for planetary missions. The quad-polarity design (Sections \ref{sec:3} and \ref{sec:4}) provides data sufficient to measure the 2 $\times$ 2 complex scattering matrix of the backscattered field, which in turn leads to the extraction of all information available in the radar reflections. The fully polarimetric radar scattering theory contains all the scattering information for any arbitrary polarization state (e.g., circular, linear and hybrid modes), thus providing better tools than CPR alone to develop scattering models of the lunar terrain \citep{Fa_2011, Campbell_2012}. Moreover, measuring the full scattering matrix allows building up a powerful observation space sensitive to shape, orientation and dielectric properties of the scatterers within a resolution cell \citep{Henderson_1998}.
\section{DFSAR Instrument overview, operations and calibration}\label{sec:3}
The DFSAR instrument is designed with a configuration that enables operations in dual-frequency, quad- and compact-polarity imaging modes to collect information about the scattering properties of the lunar surface at multiple look-angle geometry. For a complete review of technical specifications, system design and configuration of the DFSAR instrument, the reader is referred to \citet{Putrevu_2016} and \citet{Putrevu_2020}. In this section, we briefly describe the system configuration, followed by calibration of DFSAR data products.
\subsection{DFSAR Instrument Overview}\label{sec:3.1}
DFSAR is designed as two independent radar systems in L-band and S-band sharing a common microstrip planar antenna. Such a configuration enables standalone L/S-band imaging as well as synchronous L \& S imaging; the latter is ensured by having synchronization and timing signals sourced from L-band systems to that of S-band. Linear frequency modulated (LFM, or more commonly referred to as chirp) signals of requisite bandwidth (BW, selectable from 75MHz to 2MHz) is transmitted through two transmit chains feeding to H \& V ports of the antenna. Backscattered signals from the target are collected by the antenna and passed through H \& V chains of the receiver, followed by digitization, data processing and formatting in data-acquisition system to meet the data-rate constraints of the Chandrayaan-2 mission. Transmitters of each band are configured in two chains for H \& V, with Gallium Nitride (GaN) based solid-state power amplifiers (SSPAs) in high efficiency design. Each chain of SSPA has a 6-bit digital phase-shifter to enable requisite phase-setting for transmit in either circular or linear polarization. Receivers are configured with low-noise and high gain amplifiers to meet the high SNR requirements of the system. A key feature of the DFSAR system is its capability of onboard range compression implemented in data-acquisition system and provides data-rate reduction to the tune of 70\%. This compression technique is a major factor in realizing FP mode of operations, which is data-rate and volume intensive (about twice that of a hybrid-polarization implemented in Mini-RF/Mini-SAR instruments). Table \ref{Table1} lists the major specifications of the DFSAR system.

\begin{table}[htbp] 
\caption{DFSAR top-level specifications} \label{Table1}
\centering
 \begin{tabular}{|c|c|c|} 
  \hline \hline
 \textbf{Parameters} & \textbf{L-band} & \textbf{S-band} \\ \hline
  Altitude & \multicolumn{2}{c|}{Nominal 100 km} \\ \hline
 Frequency & 1.25 GHz & 2.5 GHz\\ \hline
 SAR modes & \multicolumn{2}{c|}{Single/Dual/Hybrid and Full-Polarimetry} \\ \hline
 Range swath & \multicolumn{2}{c|}{10 km} \\ \hline
 Incidence angle & \multicolumn{2}{c|}{9.6$^{\circ}$--36.9$^{\circ}$} \\ \hline
 Resolution & \multicolumn{2}{c|}{2m--75m} \\ \hline
 Chirp Bandwidth & \multicolumn{2}{c|}{75MHz--2MHz} \\ \hline
 Antenna & \multicolumn{2}{c|}{Microstrip antenna, 1.4m $\times$ 1.1m} \\ \hline
 Antenna gain & 22 dBi & 25 dBi\\ \hline
 Receiver gain & \multicolumn{2}{c|}{90 dB} \\ \hline
 Axial Ratio in Hybrid Pol mode & 0.4 dB & 1.1 dB\\ \hline
 Cross-polarization & \multicolumn{2}{c|}{$>$ 30 dB} \\ \hline
 Transmit Pulse width & \multicolumn{2}{c|}{80 $\mu$s -- 25 $\mu$s (Hybrid Pol)} \\
                      & \multicolumn{2}{c|}{50 $\mu$s -- 25 $\mu$s (Full Pol)} \\ \hline
 SSPA Peak power & 45 W & 40 W\\ \hline
 Max Duty cycle & \multicolumn{2}{c|}{24\%} \\ \hline
 Receiver noise figure & 2.8 dB & 4.3 dB\\ \hline
 Receiver front-end losses & 2 dB & 3.2 dB\\ \hline
 NESZ at swath end for 75m & -33.7 dB (Hybrid Pol) & -26.2 dB (Hybrid Pol)\\ 
 resolution and 30$^{\circ}$ incidence & -30 dB (Full Pol) & -22.5 dB (Full Pol)\\ \hline
 Onboard Processing & \multicolumn{2}{c|}{Range compression and BAQ} \\ \hline
 Data Rate & \multicolumn{2}{c|}{160 Mbps} \\ \hline
 Raw bus power & \multicolumn{2}{c|}{100 W} \\ \hline
 Payload Mass & \multicolumn{2}{c|}{20 kg} \\ \hline
   \end{tabular}
\end{table}
Onboard calibration pulses flank the imaging pulses (data-windows) and include chirp replica and noise pulses as pre- and post-imaging sequences. The replica signal refers to the signal that is coupled (-20dB) at both ends of the transmitter and the receiver. The DFSAR instrument is configured on the Chandrayaan-2 orbiter with the planar antenna mounted on the moon-viewing panel (positive Yaw) of the spacecraft cuboid, with its electronics packages integrated behind this panel. During lunar imaging, the side-looking geometry of the radar is affected by roll tilting the orbiter by the required look angle.
\subsection{DFSAR Pre-flight (Lab) Characterization}\label{sec:3.2}
\begin{figure}[htbp]
\begin{center}
\includegraphics[scale=1.3,angle=0]{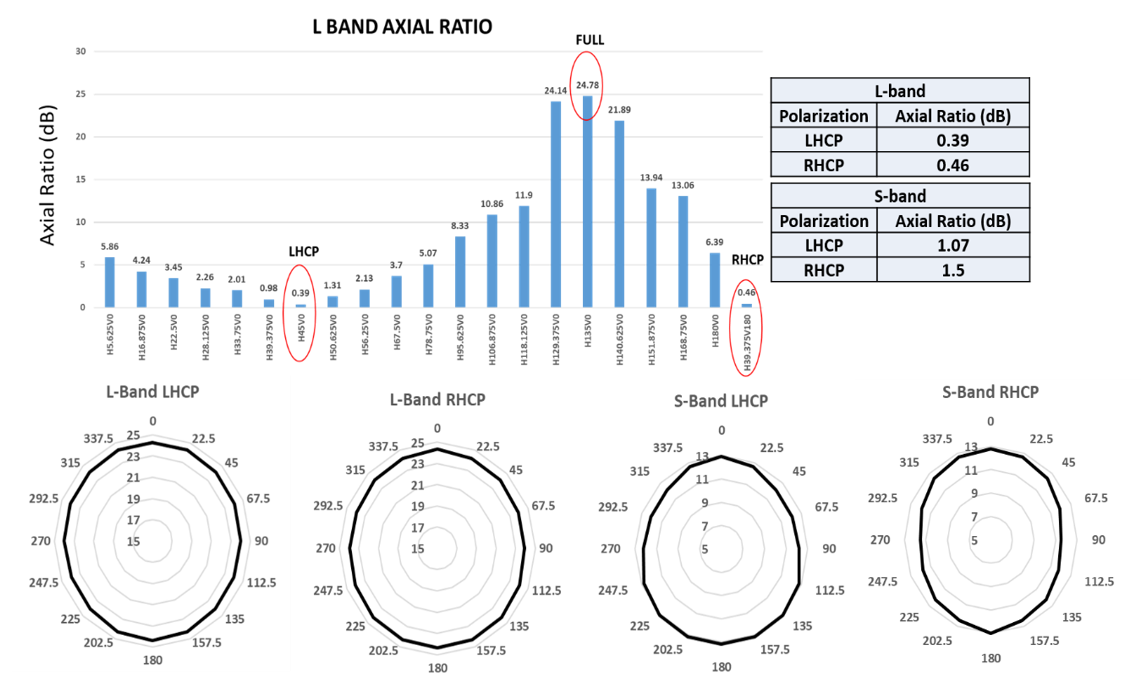}
\caption{Axial ratio measurements for various DFSAR modes of operation. \textit{Top}: The bar chart on the left illustrates ellipticity of the transmitted signals (for L-band SAR, shown as an example) corresponding to different phase-shifter settings and the inset table on the right shows axial ratio measurement results for L- and S-band transmission modes. The minimum axial ratio of 0.39 dB corresponds to phase-shifter settings of 45$^{\circ}$ and 0$^{\circ}$ of H and V transmit chains respectively, summarily represented as H45V0 on the x-axis. \textit{Bottom}: Sense of circularly polarized signals (LHCP/RHCP) is cross-verified using helical antennas of corresponding polarizations. The ellipses at the bottom of the figure represent variation of transmit signal amplitude radiated by the SAR antenna and measured using a linearly-polarized probe at different orientation angles as specified in the plots.} \label{fig:1}
\end{center}
\end{figure}

Lab characterization is essential for any instrument to meet the performance specifications, more so in the case of a polarimetric imaging system like the DFSAR. The instrument has undergone extensive in-lab characterization to ensure high quality performance in terms of radiometry as well as polarimetry. Some of the salient tests conducted are as follows:
\begin{enumerate}

    \item Radio Frequency (RF) measurements of transmit peak-power, receiver gain and noise-figure at different manual gain control (MGC) settings, front-end losses, receiver flatness and imbalances in amplitude and phase, are some of the key parameters for DFSAR characterization. These parameters have been characterized over the operating temperature range of 0$^{\circ}$ to 45$^{\circ}$ Celsius, at individual subsystem/component level. Important RF parameters were additionally characterized during thermo-vacuum tests of integrated payload.
    \item Receive chains have been characterized to generate mapping of the digitized signal counts to input signal power, which is an important input for radiometric calibration. The entire chain from input antenna port through front end elements and receiver is characterized as an integrated unit to yield the above mapping. Similarly, transmit replica calibration path is also characterized which serves as a reference to track any drifts in transmit \& receive gain.
    \item Far-field measurements of antenna patterns for both bands and H \& V polarizations have been done in a Compact Antenna Test Facility (CATF). Important parameters derived from the patterns include antenna gain, beam squint angles, cross-polarization ratio, which play an important role in establishing the calibration equation and for other post-processing requirements.
    \end{enumerate}
The above measurements meet the design specifications of the instrument and directly feed to the radiometric calibration of the SAR data, as described in section \ref{sec:3.4.1}. The received counts are first converted into expected backscattered signal power at antenna-port input and used further to derive sigma-naught ($\sigma^{\circ}$), by applying the above-mentioned lab-measurements. Sigma-naught ($\sigma^{\circ}$) is the normalized measure of the radar return from a distributed target and is defined as per unit area on the ground.    
\begin{enumerate}[resume]

    \item The DFSAR instrument is designed to operate in any of these modes: single (HH/VV), dual (HH+HV/VV+VH), hybrid (LHCP/RHCP) or Full-pol (FP). However, FP is the mode being used for systematic acquisitions of the lunar polar regions. At select locations of the Moon, imaging using hybrid-polarimetry is also being performed, for which LHCP (Left Hand Circular Polarity on transmission) mode is the default setting. For polarimetric characterization of the transmit chains, axial ratio measurements are crucial. In addition to characterizing the quality of circularly polarized transmission using axial ratios (Figure \ref{fig:17} in supporting information), the measurements have yielded the best phase settings of the 6-bit phase shifters for Left hand Circular Polarity (LHCP), Right hand Circular Polarity (RHCP) and FP modes (Figure \ref{fig:1}). 
    \item Orientation of Left/Right Circular transmit polarization is determined using helical antennas of corresponding polarizations. This has been used as a fool proof method to verify the handedness of the circular polarization.
    \item Amplitude and phase-imbalances between H \& V receive chains (covering entire path of antenna, front-end and respective receiver) for polarimetric performance.
    \item Using an external optical-fibre delay-line, end-to-end characterization of the payload was carried out. The external set-up consisted of a horn-antenna followed by a circulator, delay-line and appropriate attenuator pads; the transmit signal is collected by the horn-antenna, passed through the circulator to the delay-line; the delayed signal is fed back to the circulator, to enable its transmission towards the DFSAR antenna. Using this, the process of replica acquisition and chirp compression using on-board range compression was validated.
    \item The performance of the onboard range compression was validated using simulated distributed target data. The range-expanded data was played-back from Arbitrary Waveform Generator (AWG) to the DFSAR data-acquisition system for range compression using chirp replica from pre-imaging calibration. The onboard range-compressed data was evaluated for impulse response parameters with respect to that of conventional on-ground processed output and found to be meeting the performance requirements.
    \item Another key characterization activity is ensuring the phase synchronization between L- and S-bands to be better than 6$^{\circ}$. Chirp replicas for L- and S-band compensate for any initial bias in phases, thereby ensuring a phase-synchronization of better than 6$^{\circ}$, as measured and verified.
    \item DFSAR has an experimental radiometer mode for which the requisite measurements, data acquisition and analysis will be described in a separate paper. However, for the sake of completion, an important characterization deserving mention here is that of radiometer mode full-chain “hot-load” calibration performed using microwave absorbers. This calibration complements the “cold-sky” measurement in operational on board conditions towards retrieval of brightness temperatures.
    \end{enumerate}
 \subsection{DFSAR Operations, Data Acquisition and Products Generation} \label{sec:3.3} 
 Chandrayaan-2 was launched using ISRO’s Geosynchronous Satellite Launch Vehicle (GSLV)-Mark III on 22 July 2019 from Sriharikota, the spaceport of India. It was inserted into lunar orbit on 20 Aug, 2019 and was brought to a near-circular orbit of 80 km $\times$ 120 km, with the periapsis and apoapsis close to lunar South pole and North pole respectively. Subsequent to the initial commissioning phase activities of DFSAR during September 2019, the instrument has been tasked for systematic coverage of the lunar poles (85$^{\circ}$ and pole-wards) and is making full-polarimetric measurements in the designated time-slots (referred to as dawn-dusk seasons). Since then, DFSAR has been operational and beaming high quality data to cater to various scientific studies. After this initial campaign, DFSAR will observe the lunar polar regions ($\pm$70$^\circ$ to $\pm$90$^\circ$) at L-band along with selected non-polar targets at L- and/or S-bands. Based on this coverage, selected regions (e.g., where anomalous CPR values exist) will be observed with L- and S-band simultaneous, hybrid- and full-polarization imaging. Over the course of the mission, we will plan to acquire global L-band FP data of the lunar surface to integrate with other global lunar data sets. Raw data from the DFSAR acquisitions are downlinked at different ground stations and processed to generate products at various levels, which are denoted below:\\
\textit{Level-0A/0B}: represent range-uncompressed and –compressed, respectively. These are raw-data products with ancillary data for further processing. In this, different processes (like Block Adaptive Quantization) applied to reduce the data-rate are reversed to represent the original instrument data.\\
\textit{Level-1A}: represents seleno- tagged single-look complex (SLC) data\\
\textit{Level-1B}: represents seleno- tagged ground-range product\\
\textit{Level-2}: represents seleno-referenced product\\
A Range-Doppler algorithm \citep{Cumming_2005} is used for processing DFSAR stripmap data, which efficiently processes the SAR data in the frequency domain. The time domain correlation operation is replaced by the frequency domain multiplication operation. Each pixel of a given slant image is tagged to a particular seleno-location, i.e. latitude/longitude value using the slant range grid generated from the orbit attitude information, a digital elevation model (DEM) available at $\sim$ 118.5 meters spacing from the LRO Laser Orbital Laser Altimeter (LOLA, \cite{Mazarico_2011}) and radar parameters.
\subsubsection{Image quality analysis} \label{sec:3.3.1}
\begin{figure}[ht!]
\begin{center}
\includegraphics[scale=2]{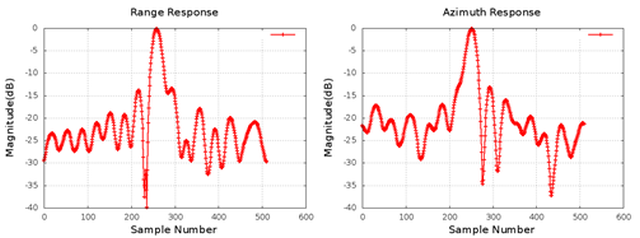}
\caption{L–Band Range (left) and Azimuth (right) Impulse response for opportunity point targets.} \label{fig:2}
\end{center}
\end{figure}

Any feature or target of dimensions similar to or smaller than that of SAR resolution cell and with sufficient backscatter energy will appear as a ``point target" in a SAR image. Such point targets are suitable candidates to characterize impulse-response of the SAR system. DFSAR data obtained during the commissioning phase has been processed and evaluated for payload performance. In this process, range and azimuth impulse responses have been analysed for opportunity point targets, since no ideal point targets are available on the lunar surface. System impulse response characteristics such as peak-to-sidelobe ratio (PSLR) and resolution have been derived from these opportunity targets and some of the sample results are shown in Table \ref{Table2} and Figure \ref{fig:2}. After the radiometric calibration process (described in section \ref{sec:3.4.1}) of DFSAR, noise-equivalent sigma-naught (NESZ) has been derived (Table \ref{Table3}) from the noise-floor of the system, measured during on-board calibration data acquisition. NESZ represents the lowest $\sigma^{\circ}$ that can be measured by the instrument. Radiometric resolution has been estimated over homogeneous regions located in the processed images. All the DFSAR performance metrics are found to be matching the pre-launch expected values.\\
\begin{table}[h!] 
\caption{Impulse response performance obtained from DFSAR data acquired in different polarimetric modes} \label{Table2}
\centering
 \begin{tabular}{|c| c| c| c| c| c|} 
 \hline\hline
 Mode & Chirp BW & Range PSLR & Azimuth PSLR & Range & Azimuth \\ 
  & (MHz) & (dB) & (dB) & resolution (m) & resolution (m) \\
\hline
L band FP & 7.5 & -11.7 & -10.9 & 18 & 12.8\\
L band CP & 75 & -15.35 & -15.56 & 1.9 & 2.3\\
S band, CP & 50 & -13.9 & -13.24 & 2.7 & 3.03\\
\hline
 \end{tabular}
\end{table}

\begin{table}[ht!] 
\caption{Noise-equivalent sigma-naught (NESZ) performance metrics for DFSAR polarimetric modes using SAR noise calibration data, with signal power normalized to 100km altitude, and at 30$^{\circ}$ angle of incidence.}
\label{Table3}
\centering
 \begin{tabular}{|c| c| c | c|} 
 \hline\hline
 Mode & Chirp Bandwidth & pre-launch NESZ estimate & post-launch NESZ value \\ [0.5ex] 
& (MHz) & (dB) & (dB) \\ \hline
L-band FP & 7.5 & -27.7 & -27.9\\
L-band CP & 7.5 & -29.5 &-29.8\\
S-band FP & 7.5 & -22.6 &-23.8\\
S-band CP & 7.5 & -24.3 & -25.6\\
\hline
 \end{tabular}
\end{table}

\subsection{Calibration of the data-products}\label{sec:3.4}
As the DFSAR is a polarimetric system with capability of acquiring quad-pol and compact polarimetric (CP) data, radiometric and polarimetric calibration are critical for the derivation of lunar surface physical parameters.
\subsubsection{Radiometric Calibration}   \label{sec:3.4.1}
\begin{figure}[htbp]
\begin{center}
\includegraphics[scale=2]{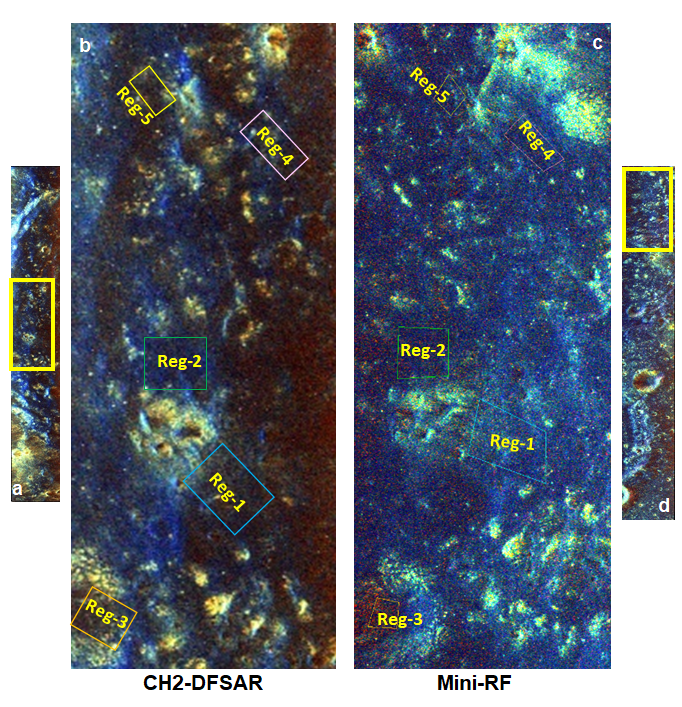}
\caption{Radiometry comparison over five sample regions (indicated by boxes in (b) and (c)) between (a) S-band DFSAR and (d) S-band Mini-RF hybrid-polarimetric data shown here in \textit{m-chi} decomposition images with the RGB colour scheme indicating the dominant scattering regime: even bounce (red); odd bounce (blue); and volume scattering (green). The extent of zoomed-in portions (b) and (c) are shown with yellow boxes in (a) and (d) respectively. Both DFSAR and Mini-RF images shown here are resampled to similar spatial resolution ($\sim$ 30 m/pixel) and the centre coordinates of DFSAR image in (a) are 35.36$^{\circ}$N, 79.54$^{\circ}$E} \label{fig:3}
\end{center}
\end{figure}

Radiometric calibration provides the coefficients to map SAR data (pixel values) to their corresponding backscatter coefficients, i.e., the normalized radar cross section (NCRS) of targets. Due to the absence of any calibration targets on the Moon with well-known RCS in L\&S bands, the calibration is performed based on the radar-equation, using lab characterized system parameters.
\begin{equation} \label{eq:1}
P_r = k_{v} \times(DN)^{2}\times g_{oa}\times g_{or}\times \frac{L_{b}}{SF}
\end{equation}   
where, 
DN represents the digital number corresponding to a pixel after processing; k$_{v}$ is a factor that
accounts for quantization to convert count to power at the input of analog-to-digital converter
(ADC) with 50 ohm impedance; g$_{oa}$ and g$_{or}$ are the Pulse Repetition Frequency (PRF) / Doppler bandwidth and the range sampling frequency / range bandwidth corresponding to the oversampling factors for azimuth and range respectively; L$_{b}$ is beam shape loss; and SF is scale factor applied during onboard processing, for a given polarization chain.\\
Using the above equation, processor gain related factors are first corrected to convert the digital
counts into raw power, as would be present at the receiver output. The raw power as computed
from \ref{eq:1} is then inverted using \ref{eq:2} to derive sigma-naught ($\sigma^{\circ}$), with the inputs from lab-measured parameters.
\begin{equation} \label{eq:2}
P_r = \frac{P_{t}G_{t}G_{r}\lambda^{3}\sigma^{\circ}}{(4\pi)^{3}R^{3}LL_{S}}\times \frac{c\tau}{2\sin\theta}\times \frac{G_{rx}}{MGC_{rx}}
\end{equation}
where, \\
P$_{r}$ is power received corresponding to a digital-count of a pixel;
P$_{t}$ is transmitted power;
G$_{t}$ and G$_{r}$ are transmit and receive antenna gains respectively;
G$_{rx}$ is the receiver gain; R is the slant-range of the pixel location; $\lambda$ is radar wavelength, $\theta$ denotes incidence angle, and c is velocity of light; $\tau$ is pulse width or chirp duration;
MGC$_{rx}$ is attenuation setting by the manual-gain controller of the receiver; L$_{s}$ is receiver chain front end loss, and L is azimuth antenna length.\\
The processed images are corrected using the lab-measured antenna patterns for radiometric normalization, so that a uniform / single calibration equation can be used for the entire dataset, irrespective of the pixel position. In the case of Level-1A (SLC) products, the relationship between the digital number (DN) and the backscattering coefficient $\sigma_\theta$ for each polarization channel can be written as:
\begin{equation} \label{eq:3}
    \sigma^{\circ}_{i,j} = \frac{DN_{i,j}}{10^{(\frac{K}{10})}}; \quad i=1,2\dots L; j=1,2\dots M
\end{equation}
Where DN = I+Q\textit{i}, and I and Q are respectively the real part and the imaginary part of the complex pixel value; K is the absolute calibration constant, and is constant over the entire SLC products; and L, M are the number of lines and columns in the product. The calibration constant (K) is provided with the associated xml label file for each data set and is different for products Level-1A and Level-1B and above. For Level-1B and above products, $\sigma^\circ$ in linear scale to dB can be directly derived as:
\begin{equation} \label{eq:4}
    \sigma^{\theta}[dB] = 20\log_{10}(DN)-K 
\end{equation}
Specimen results as shown in Figure \ref{fig:3} and Table \ref{Table4} compares radiometry of S-band DFSAR hybrid-polarimetric mode data with S-band Mini-RF data over a part of Gauss crater region that has overlap from both the instruments. The radiometric accuracies (with incidence angle independent gamma-naught ($\gamma^\circ$) obtained over different sample regions are found to be better than 2dB. In the absence of any L-band data nor any full-polarimetry data of the Moon, validation could only be done for the process adopted for radiometric calibration, using S-band hybrid-polarity data of Mini-RF as reference. Considering that there are no additional processes which may hinder such a generalization for either L-band or full-polarimetry data, radiometric accuracy of better than 2dB can be safely considered as a worst-case value.\\

\begin{table}[ht] 
\caption{Radiometry ($\gamma^\circ$) comparison of S-band DFSAR with Mini-RF data over five selected regions as shown in figure \ref{fig:3}. LH and LV stands for Left circular transmit and Horizontal (H), Vertical (V) receive polarizations respectively.} 
\label{Table4}
\centering
 \begin{tabular}{|c|c|c|c|c|c|c|} 
  \hline \hline
 \textbf{Region} & \multicolumn{2}{c|}{\textbf{DFSAR S-band CP}} & \multicolumn{2}{c|}{\textbf{Mini-RF S-band CP}} & \multicolumn{2}{c|}{\textbf{$\gamma^{\circ}$}}\\ 
 & \multicolumn{2}{c|}{$\gamma^{\circ}$ (dB)} & \multicolumn{2}{c|}{$\gamma^{\circ}$ (dB)} & \multicolumn{2}{c|}{\textbf{difference (dB)}} \\ \cline{2-7}
 &  LH & LV & LH & LV & LH & LV\\ \hline
 Reg--1 & -13.78 & -13.60 & -12.49 & -12.54 & -1.29 &-1.06 \\ \hline
 Reg--2 &-13.18 &-13.06 & -14.80 &-14.64 &1.62 &1.58 \\ \hline
 Reg--3 &-15.19 &-15.14 &-15.88 &-15.65 &0.69 &0.51 \\ \hline
 Reg--4 &-12.99 &-12.96 &-13.58 &-13.69 &0.59 &0.73 \\ \hline
 Reg--5 &-14.79 &-14.28 &-14.08 &-14.45 &-0.71 &0.17 \\ \hline
   \end{tabular}
\end{table}

\subsubsection{Polarimetric Calibration for Full-Polarimetry} \label{sec:3.4.2}
Polarimetric characterization of a region of interest is heavily dependent on the quality of polarimetric calibration that establishes the inter-channel (HH/VV/HV/VH) phase-relationships. Calibration errors can affect polarimetry products (coherency \& covariance matrices and the other derived daughter parameters) thereby leading to misinterpretation of scattering/physical processes under study. The “measured” scattering matrix (of a target) is a product of “actual” scattering matrix (section \ref{sec:4.1}) and the distortion matrices corresponding to transmit [T] and receive [R] paths of the radar signal. Considering the order of interactions between the signal and target, the “measured” scattering matrix may be written as:
\begin{equation} \label{eq:5}
    \begin{bmatrix}
    S_{HH}& S_{HV} \\
    S_{VH} & S_{VV} 
\end{bmatrix}_{measured} = 
    K(\gamma)[R]
        \begin{bmatrix}
        S_{HH}& S_{HV} \\
        S_{VH}& S_{VV} 
        \end{bmatrix}_{actual} [T]
\end{equation}    
with \\
\begin{align*}
    [R] = \begin{bmatrix}
                    f_{r}(\gamma)& \delta_{1}^{r}(\gamma) \\
                    \delta_{2}^{r}(\gamma)& 1 
                    \end{bmatrix}\\
    [T]= \begin{bmatrix}
                    f_{t}(\gamma)& \delta_{1}^{t}(\gamma) \\
                    \delta_{2}^{t}(\gamma)& 1 
                    \end{bmatrix}                
\end{align*} 
Where f$_{t}$ and f$_{r}$ represent channel imbalances in HH polarization with reference to that of VV, for transmit and receive paths respectively. They comprise of differences in antenna patterns between the polarizations and also the gain and phase of their respective paths. $\delta_{1}$ and $\delta_{2}$ represent the cross-talk terms HV and VH respectively, mainly arising from the antenna cross- polarization components. As we are dealing with imaging in lunar conditions with negligible ionosphere, Faraday rotation of the signals is not expected and thereby not considered in the above equation. After the SAR image formation and radiometric calibration, polarimetric calibration is applied to the resulting data. Methodology adopted for this calibration closely follows that of \citet{Sun_2018} due to the following major factors: 1) the methodology does not depend on any external calibration targets (like corner reflectors) and is derived from the SAR data itself, which is a necessity in the case of lunar imaging. 2) Cross-talk estimation and channel-imbalances are estimated in an iterative fashion due to their mutual dependency 3) The computations are based on derived covariance matrix, which results in more limiting equations to derive the phase-estimates. 
\begin{figure}[htbp]
\begin{center}
\includegraphics[scale=1.5]{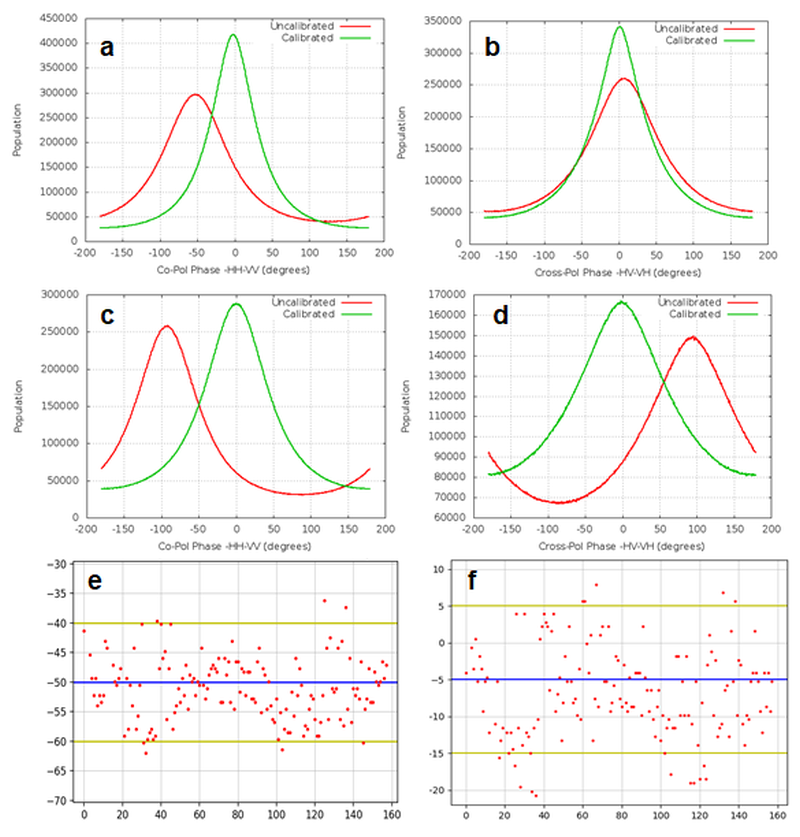}
\caption{DFSAR polarimetric calibration results. (a), (b) histograms of L-band co-pol (HH/VV) and
cross-pol (HV/VH) phase differences; (c), (d) histograms of S-band co-pol (HH/VV) and cross-pol
(HV/VH) phase differences; (e), (f) stability of co- and cross-pol phase differences across several L-
band datasets, prior to applying the correction (i.e., uncalibrated) respectively. This illustrates that a single set of correction factors are uniformly applicable across the acquired datasets.} \label{fig:4}
\end{center}
\end{figure}
DFSAR polarimetric calibration is based on the above methodology while focusing on dominant surface and volume backscatter regions for co- and cross-polarization signal responses. The results in figure \ref{fig:4} show phase histograms for uncalibrated and calibrated L- band (figure \ref{fig:4}a, \ref{fig:4}b) and S-band (figure \ref{fig:4}c, \ref{fig:4}d) acquisition, corresponding to co- and cross- polarization cases. Phase correction of the order of $-50^{\circ}$ for co-polarization phase and $-5^{\circ}$ for cross-polarization phase are required for phase calibration of DFSAR L-Band acquisitions. Corresponding numbers for calibration of S-band FP acquisitions are $-100^{\circ}$ and $+100^{\circ}$ for co- and cross-polarization phase respectively. For L-band data, the stability of the co- and cross- polarization phases observed over multiple acquisitions (figure \ref{fig:4}e, \ref{fig:4}f) is typically within $\pm10^{\circ}$, indicating that a uniform set of polarimetric calibration factors can be applied for the DFSAR datasets. However, monitoring of polarimetric calibration would be continued for future acquisitions and any deviations will be taken care of by estimating new calibration coefficients and incorporating them during the subsequent processing of the DFSAR data.   
    
\begin{figure}[htbp]
\begin{center}
\includegraphics[scale=2.2]{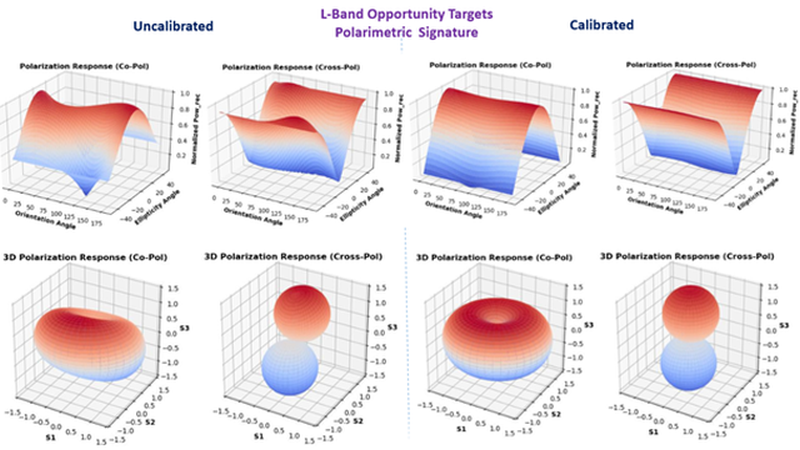}
\caption{Uncalibrated (left set) and calibrated (right set) Polarimetric signatures corresponding to an opportunity point target, to assess the efficacy of polarimetric calibration. The response for each Stokes parameter (S1, S2, S3) closely match with those expected from an ideal trihedral corner reflector, thereby validating the polarimetric calibration exercise.} \label{fig:5}
\end{center}
\end{figure}

For a holistic view of the quality of polarimetric calibration, opportunity point targets have been analysed for polarimetric signatures, both in terms of polar coordinates (ellipticity angle and orientation angle) and Cartesian coordinates representing Stokes parameters \citep{Raney_2007, LeeJong-Sen2009} over a Poincaré sphere. The results (Figure \ref{fig:5}) show an excellent match with that of an ideal trihedral corner reflector \citep{van_Zyl_1987}, and give the required confidence on the quality of polarimetric calibration.   
\FloatBarrier
\subsubsection{Polarimetric Calibration for Compact-Polarimetry} \label{sec:3.4.3}
\begin{figure}[htbp]
\begin{center}
\includegraphics[scale=2]{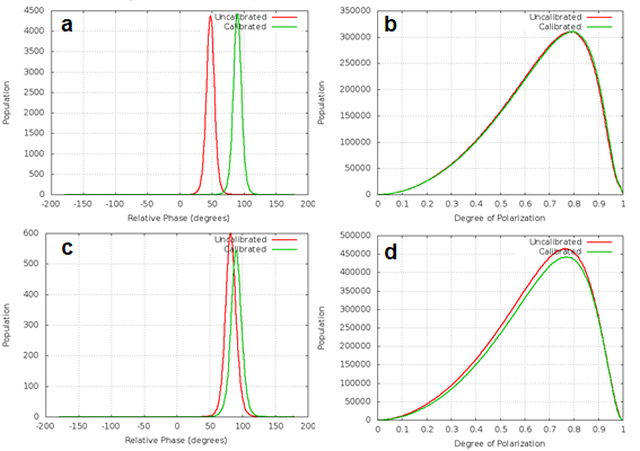}
\caption{Relative phase ($\delta$) and degree of polarization (\textit{m}) histograms for DFSAR (a), (b) L-band CP and (c), (d) S-band CP data respectively after applying the necessary calibration factors.} \label{fig:6}
\end{center}
\end{figure}

The calibration approach for CP data is a subset of that adopted for calibration of full-polarimetric data. The channel and phase imbalances were derived from L- and S-band data over some uniform regions on the lunar surface. The relative phase histograms computed from Stokes parameters \citep{Raney_2007, Raney_2011} were derived over several CP data acquisitions to establish the consistency of amplitude and phase calibration parameters. The order of relative phase ($\delta$) corrections is about 40$^{\circ}$ for L-band and 10$^{\circ}$ for S-band data to achieve the phase calibration. The results of applying this correction is shown in Figure \ref{fig:6} in the form of uncalibrated and calibrated $\delta$ and degree of polarization (\textit{m}) histograms for a set of DFSAR L- and S- band CP data.  
\FloatBarrier
\section{DFSAR data analysis and Interpretation}\label{sec:4}
\subsection{Quad-polarity data parameters} \label{sec:4.1}
In FP mode, the DFSAR transmits two orthogonal polarizations (Horizontal ‘H’ and Vertical ‘V’) on a pulse-to-pulse basis and receives the scattered waves in two orthogonal polarizations (in the same basis as used for transmission). Thus, for one snapshot of the received echoes, we can construct a 2$\times$2 complex scattering matrix [S] that relates the two-dimensional transmitted and received electric field vectors [E] as \citep{LeeJong-Sen2009, Cloude_2009}:
\begin{equation} \label{eq:6}
\begin{bmatrix}
 E_{H}\\
 E_{V} 
\end{bmatrix}_{receive} = 
\frac{exp(-ikr)}{r}  \begin{bmatrix}
                    S_{HH}& S_{HV} \\
                    S_{VH}& S_{VV} 
                    \end{bmatrix}
                        \begin{bmatrix}
                         E_{H}\\
                         E_{V} 
                        \end{bmatrix}_{transmit}
\end{equation}
Where all the elements are complex-valued and the subscripts H and V indicate Horizontal and Vertical polarization respectively. The factor exp(-ikr)/r, where k = 2$\pi/\lambda$ is the wave number, expresses the attenuation for a spherical wave of a radius that equals the distance (r) between the scatterer and the radar. The elements of [S] describe the amplitude and phase changes imposed on the incident field by the scattering process, and in monostatic configurations like that of DFSAR, [S] becomes symmetric, i.e., S$_{HV}$ = S$_{VH}$ for all reciprocal scattering media \citep{LeeJong-Sen2009, Campbell_2002}. Once the full scattering matrix is measured, any arbitrary complex scattering amplitude can be reconstructed as a linear combination of the elements of the measured scattering matrix.

In order to describe the polarimetric scattering behaviour of the lunar surface, which is a distributed target, a second-order statistical formalism is required. The most common formalism to fully characterize distributed scatterers is the 3$\times$3 coherency [T] or covariance [C] matrix defined by the outer product of a three-dimensional scattering vector in the Pauli or lexicographic formulation (for a review, the reader is referred to \citet{LeeJong-Sen2009}). Followed by this, multilook processing (averaging several independent estimates of reflectivity) on [T] and [C] matrices is required to assess the scattering mechanism of distributed targets. This averaging process (of ‘N’ looks) also reduces the speckle (by a factor of N$^{1/2}$) and degrades the azimuth resolution (by a factor of N). As the SAR image is formed by coherently processing returns from successive pulses, this results in a pixel-to-pixel variation in intensity, and the variation manifests itself as a granular pattern called speckle \citep{LeeJong-Sen2009}. A common approach to speckle reduction is to average several independent estimates of reflectivity, as described above. After calculating the multilooked [T] and [C] matrices and by assuming that there is no correlation between the cross-polarized (HV, VH) and like-polarized (HH, VV) linear components for natural surfaces \citep{Campbell_2012}, it is straightforward to calculate the backscatter coefficients ($\sigma^{\circ}$) in the two circularly polarized components from the linearly polarized terms as given below \citep{Campbell_2002, Campbell_2012}:
\begin{equation} \label{eq:7}
Same\: sense\: Circular\: (\sigma^{\circ}_{SC}) = \frac{1}{4}\left[\sigma^{\circ}_{HH}+\sigma^{\circ}_{VV}+4\sigma^{\circ}_{HV}-2Re(S_{HH}S^{*}_{VV})\right]
\end{equation}
\begin{equation} \label{eq:8}
Opposite\: sense\: Circular\: (\sigma^{\circ}_{OC}) = \frac{1}{4}\left[\sigma^{\circ}_{HH}+\sigma^{\circ}_{VV}+2Re(S_{HH}S^{*}_{VV})\right]
\end{equation}
The CPR, which provides significant information on the physical characteristics of the target surface (e.g. \cite{Campbell_1993, Rignot_1995, Harmon_1994, Harmon_2001, Campbell_2012} can thus be calculated using:
\begin{equation} \label{eq:9}
CPR\: = \frac{\sigma^{\circ}_{SC}}{\sigma^{\circ}_{OC}}
\end{equation}
Scattering decompositions are widely applied over polarimetric radar data for interpretation, physical information extraction, segmentation and/or as a pre-processing step for geophysical parameter inversion (e.g. \cite{Cloude_1996, Cloude_1997, LeeJong-Sen2009, Raney_2012, Cloude_2012}). In general, the decompositions of second-order scattering matrices (i.e., [T] or [C]) are rendered into two classes: Eigenvector and eigenvalue based decompositions and model-based decompositions \citep{Cloude_1996}. For the lunar surface, it is of interest to generate the concept of an average or dominant scattering mechanism for the purposes of classification or inversion of scattering data through such target decomposition theorems. For DFSAR quad-pol data results shown in section \ref{sec:6.1} of this paper, we adopted two widely used decomposition methods which are the Eigenvector and Eigenvalue based \citep{Cloude_1996} and Yamaguchi 4-component decomposition \citep{Yamaguchi_2005} methods. Specifically, we derive two important statistical parameters arising directly from the eigenvalues of the coherency matrix, which are the polarimetric scattering entropy (H) and the alpha angle ($\alpha$) to extract physical information from the observed scattering of microwaves by the lunar surface. They are calculated for the backscattering case as follows \citep{Cloude_1996}:
\begin{equation} \label{eq:10}
P_i = \frac{\lambda_{i}}{\sum\limits_{j=1}^3 \lambda_{j}}
\end{equation}
\begin{equation} \label{eq:11}
H = -\sum\limits_{i=1}^3 P_{i}log_{3}P_{i}
\end{equation}
Where 0 $\leq\lambda_{3}\leq\lambda_{2}\leq\lambda_{1}$ are the real nonnegative eigenvalues of [T] and P$_{i}$ expresses the appearance probability for each contribution. For each eigenvector the roll-invariant scattering alpha angle is calculated as:
\begin{equation} \label{eq:12}
\alpha = \sum\limits_{i=1}^3 P_{i}\alpha_{i}
\end{equation}
The ranges of entropy ($0<H<1$) can be interpreted as a measure of the randomness of the scattering process. An entropy of H=0 indicates a non-depolarizing scattering process described by a single scattering matrix, while the other extreme, an entropy of H=1 indicates the presence of a random noise scattering process, which depolarizes the incident wave completely regardless of its polarization \citep{LeeJong-Sen2009}. The average alpha angle ranges between 0 and 90 degrees and is associated to the type of corresponding scattering mechanism. In general, 0$^{\circ}\leq\alpha\leq30^{\circ}$ corresponds to surface scattering processes, 40$^{\circ}\leq\alpha\leq50^{\circ}$ to dipole-like scattering behaviour and finally 60$^{\circ}\leq\alpha\leq90^{\circ}$ indicates dihedral or helix type scattering mechanisms \citep{LeeJong-Sen2009}.
\subsection{Compact-polarity data parameters} \label{sec:4.2}
While the FP SAR data mode in DFSAR provides optimum performance in target characterization due to its complete radar target information content, compact polarimetric (CP) SAR mode offers more information than acquired from single or dual-polarized SAR mode (e.g. \cite{Raney_2011}). The hybrid polarimetry mode of DFSAR follows from the Mini-SAR and Mini-RF radar configuration \citep{Nozette_2010}, which receives orthogonal linear polarizations while transmitting circular polarization. The hybrid polarimetric architecture of the DFSAR is a form of compact polarimetry \citep{Raney_2011} and the resulting data from this mode support calculation of the 2 $\times$ 2 covariance matrix of the backscattered field, from which follow the four Stokes parameters required to calculate $\sigma^{\circ}_{OC}$, $\sigma^{\circ}_{SC}$, and CPR. For brevity, we do not elaborate here the derivations of these parameters including the \textit{m-chi} decomposition technique that are used for the analysis of DFSAR CP data (for a review, the reader is referred to \citet{Raney_2012}).
\subsection{Image artifacts}
Some DFSAR images contain vertical streaks near the crater rims, with a spread in the azimuth direction. This has been observed most prominently at the far wall (from the radar's perspective) of large impact craters that perfectly align (i.e., oriented perpendicularly) with the incident radar signals, giving rise to saturation of the signal which after SAR processing appears as streaks (for e.g., see the red vertical streaks in figure \ref{fig:8}). This is similar to the ``cardinal effect" observed commonly from urban regions in terrestrial radar images due to the tendency of a radar to produce very strong echos from a city street pattern or other linear feature oriented perpendicular to the radar beam (e.g. \cite{Henderson_1998}).\\
Other artifacts in the form of irregular patches observed in some DFSAR FP images are a manifestation of azimuth ambiguities (for e.g., see the red irregular patches in figures \ref{fig:8} and \ref{fig:11} a). These are a result of the target returns that enter the system through azimuth sidelobes of the antenna pattern, and are aliased onto the processing (Doppler) bandwidth. Though the ambiguity levels are quite low and visible in only FP co-pol backscatter images (HH, VV) after stretching/scaling, these artifacts are emphasized greatly in the decomposition images. One of the ways to reduce this azimuth ambiguity is to further increase the pulse repetition frequency (PRF), which is already doubled due to FP SAR requirements. However, increasing the PRF would have challenges in accommodating the data-window and can lead to range ambiguities. Also, this would be only possible in new data acquisitions with new PRF settings. Hence, we have very recently adopted an alternative method, which is to process the (already acquired) data with reduced processing bandwidth (say, half) and thereby, excluding the aliased component of the returns. The reduced processing bandwidth impacts the single-look azimuth resolution, however, in case of the DFSAR it can be taken care of by adjusting the corresponding multi-look factor. We have tested this processing method on one of the FP datasets, and significant improvements are observed in the sense that most of these azimuth ambiguity related artefacts are removed. After testing this method over several datasets, we plan to process all the existing DFSAR FP data through this new processing chain.

\section{Data used and study area} \label{sec:5}
\begin{figure}[ht]
\begin{center}
\includegraphics[scale=0.13]{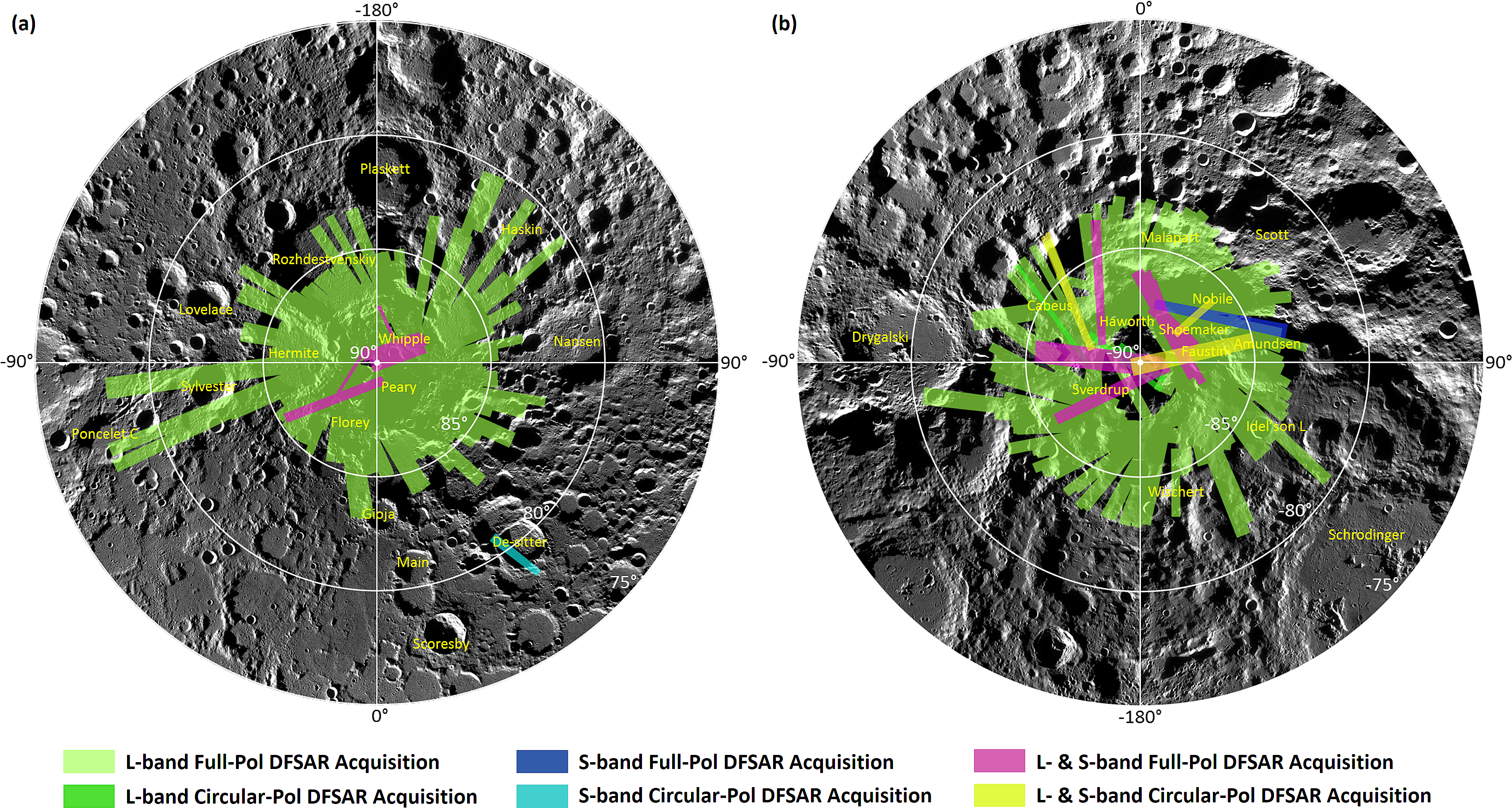}
\caption{Data Acquisition by Chandrayaan-2 DFSAR over (a) North and (b) South pole regions of the Moon. Both L- and S-band DFSAR image swaths acquired in different polarimetric modes (as indicated in the legend below) are overlaid on LROC WAC polar mosaics for context.} \label{fig:7}
\end{center}
\end{figure}

As the DFSAR system probes the lunar regolith at two frequencies (L- and S-band) it provides additional information on the physical properties of the upper meter or two of lunar surface, as we show with some examples in the following section. Here, we present the first results of a campaign designed to evaluate the potential of quad- and compact-polarimetry at L- and S-bands using the new DFSAR data. To this end, several images over polar regions ($\sim$40 \% of 80$^{\circ}$-90$^{\circ}$ latitude) and a few non-polar regions have been acquired through October 2020, providing data at different sensor configurations (polarization channels and incidence angles). Figure \ref{fig:7} shows DFSAR data coverage in both polar regions of the Moon. For this preliminary analysis, to characterize the crater ejecta and floor materials at L- and S-band wavelengths, we sampled three different regions from the polar and non-polar regions of the lunar surface. We restricted our analysis to crater interiors and their continuous ejecta blankets to isolate the response of the ejecta from surrounding materials. Typically, this (ejecta) region encompasses surface material within approximately 1-2 crater radii of the crater rim (e.g. \cite{Thompson_1981}).

We analysed L-band FP mode data of secondary craters on the floor of Peary crater (88.6$^{\circ}$N, 33$^{\circ}$E) in the north polar region, L- and S- band (dual-frequency) CP data of Byrgius C crater (21.17$^{\circ}$S, 64.52$^{\circ}$W) located between Mare Humorum to the west and the Orientale impact basin to the east, and L-band CP mode data over a region near Manzinus C crater (70$^{\circ}$S, 21.6$^{\circ}$E) in the south polar region. Since each of the above datasets are acquired at different configurations, we processed them with different numbers of looks and at different resolutions, which are explained in the following section. For each case (polarimetric mode of acquisition), we focus on comparisons between DFSAR data and S-band zoom mode \citep{Raney_2011} data obtained from the Mini-RF instrument on board LRO. For context and comparison, we utilized optical images obtained from the Lunar Reconnaissance Orbiter Camera (LROC) Wide Angle (WAC) and Narrow Angle Camera (NAC, \citet{Robinson_2010}).
\section{Results} \label{sec:6}
\subsection{L-band FP data} \label{sec:6.1}
\begin{figure}[htbp]
\begin{center}
\includegraphics[scale=0.15]{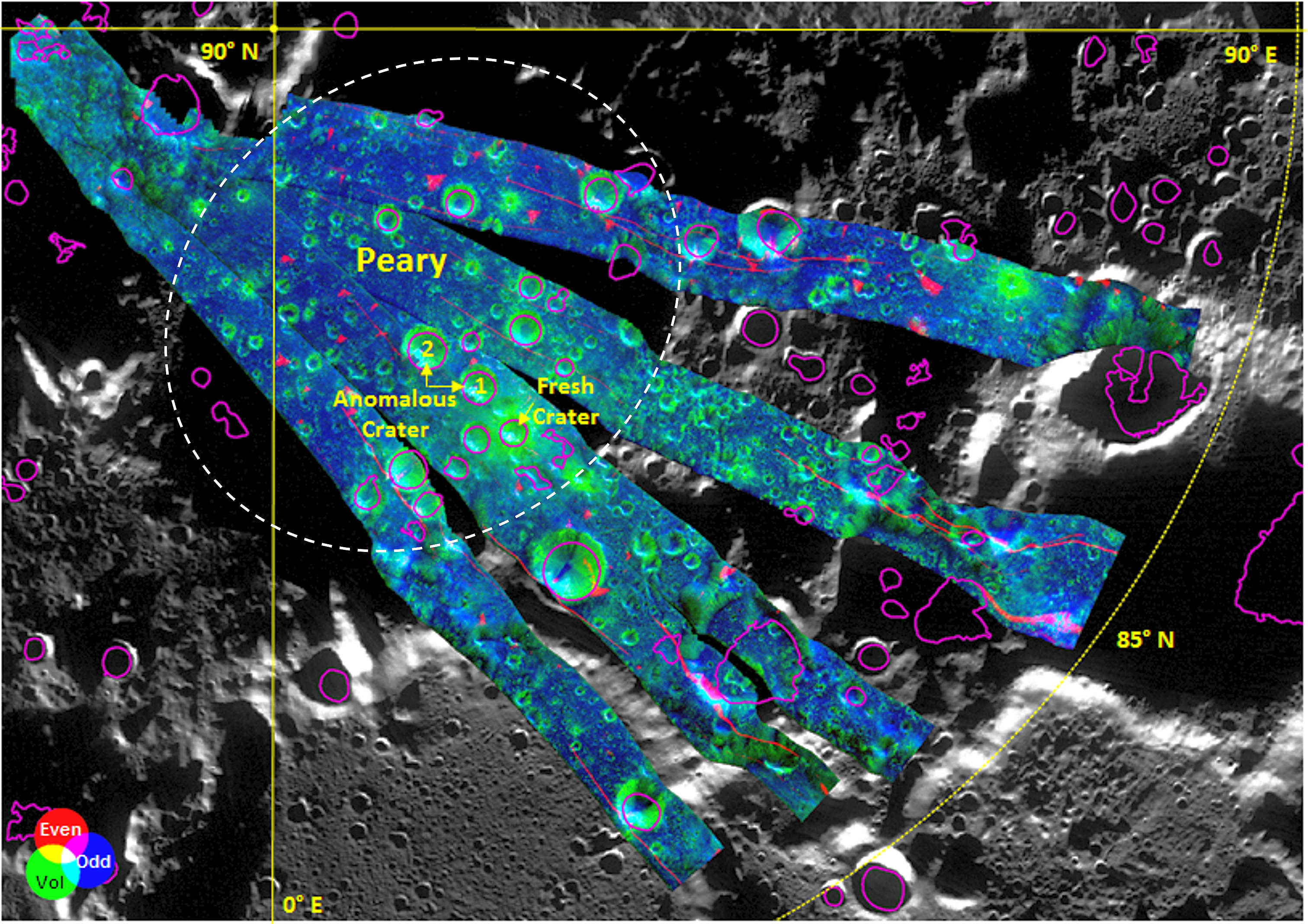}
\caption{Peary crater (78 km diameter; 88.6$^{\circ}$N, 33$^{\circ}$E) in the lunar North Pole region as viewed by L-band DFSAR in Yamaguchi 4-component decomposition images. The terrain corrected DFSAR decomposition images are mosaicked and overlain on a LROC WAC North Pole mosaic. The white dashed polygon outlines the approximate extent of Peary crater rim and the polygons in magenta are the regions of permanent shadow as obtained from LRO’s LOLA instrument. The two “anomalous” craters and a “normal” fresh crater (marked with numbers “1”, “2” and “fresh crater” respectively) analysed in this study are shown with arrows in yellow, and are also separately shown in figures \ref{fig:10} and \ref{fig:11}. The colour wheel highlights the colours for each scattering regime (red: even bounce; blue: single (odd) bounce; green: volume scattering). Note that the vertical lines and small patches in red near crater rims and edges of DFSAR image strips that appear here are artifacts as described in section \ref{sec:4} and are not to be confused with the dominant scattering mechanism from those regions.} \label{fig:8}
\end{center}
\end{figure}

\begin{figure}[htbp]
\begin{center}
\includegraphics[scale=0.13]{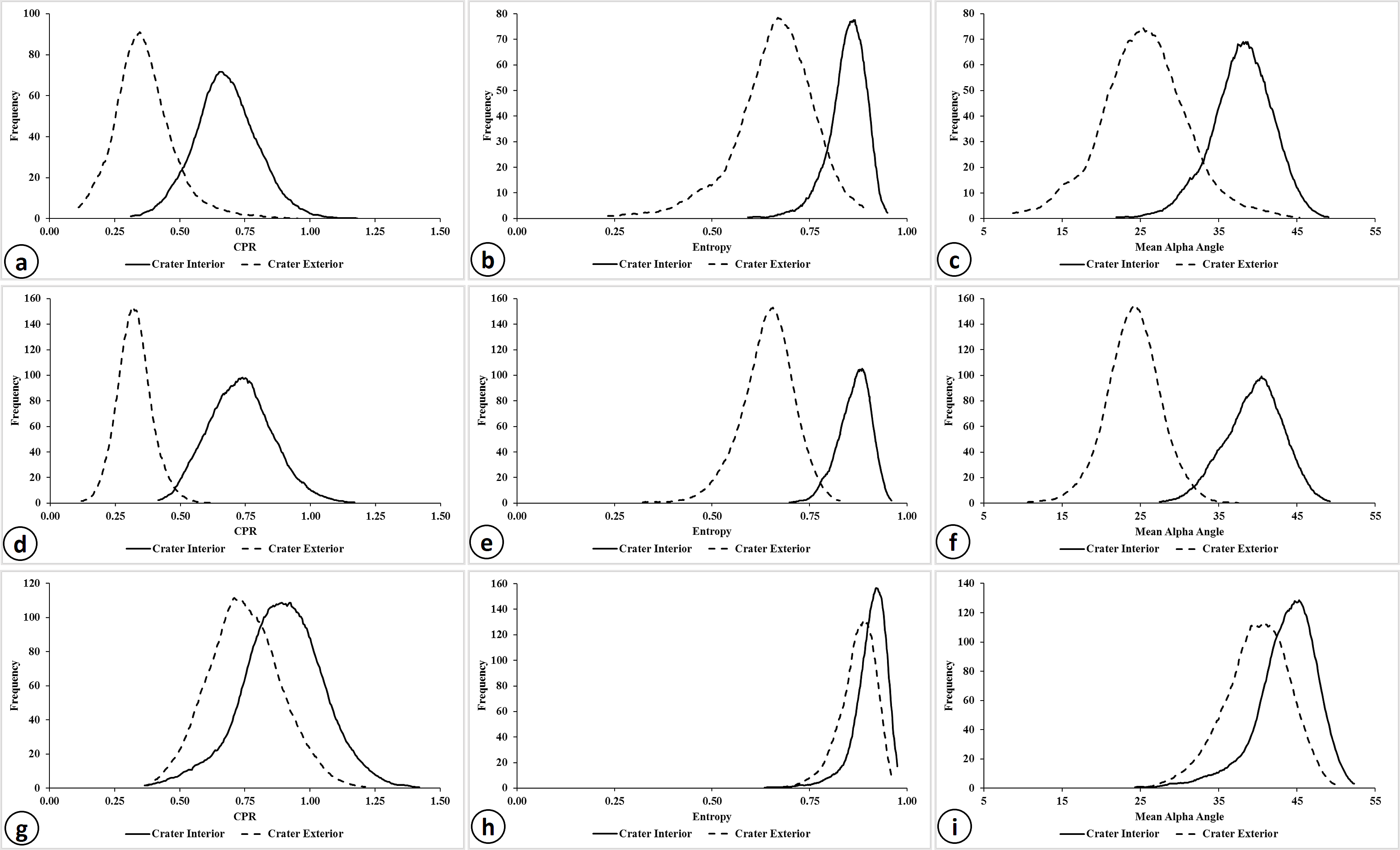}
\caption{From left to right are histograms of CPR, entropy, and mean alpha angle for interior and exterior regions of two anomalous craters “1”, “2” and a fresh crater shown in the top [(a)-(c)], middle [(d)-(f)], and bottom [(g)-(i)] rows respectively. Note the differences in the distributions of entropy and alpha angle values compared to those of CPR for each crater region. The polygons shown in the interior and exterior regions of the craters as shown in figure \ref{fig:11} b are used to generate these histograms.} \label{fig:9}
\end{center}
\end{figure}

\begin{figure}[htbp]
\begin{center}
\includegraphics[scale=1.5]{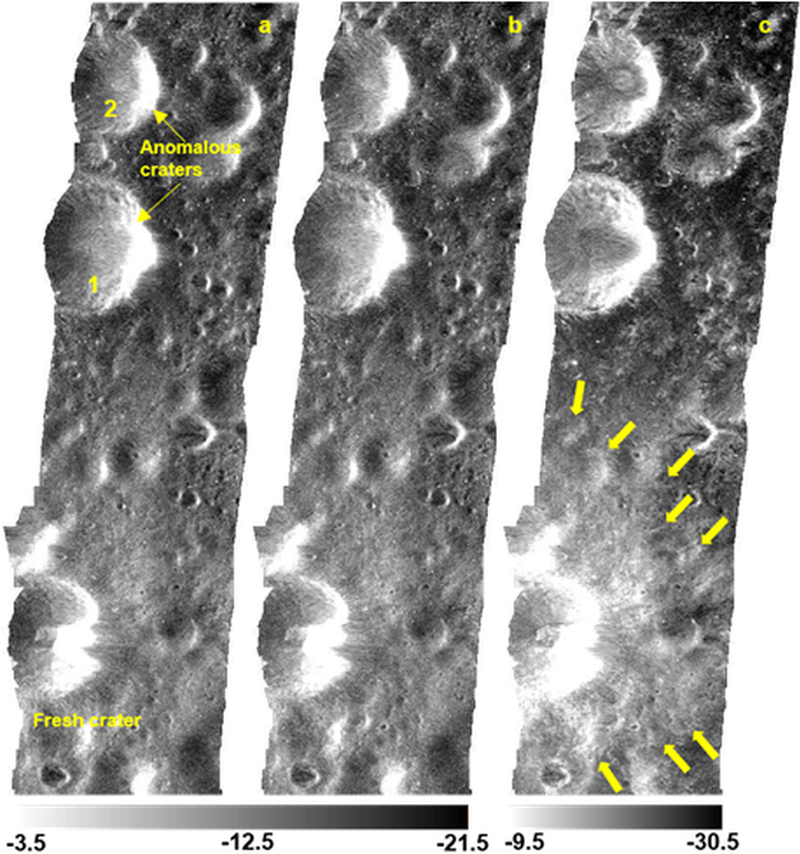}
\caption{L-band orthorectified, calibrated radar polarization images of secondary craters on the floor of Peary crater. (a) HH (b) VV and (c) HV polarization images shown in dB scale with ranges shown at the bottom. The anomalous craters (shown with arrows and marked “1”, “2”) and the young, fresh crater analysed in this work are indicated in the HH-pol image. Note the proximal ejecta region of the fresh crater that is clearly highlighted (approximate boundary indicated with yellow arrows) in the HV-pol image compared to the HH- and VV-pol images. Also apparent from these images is that the depolarized echoes (HV-pol in this case) are much more sensitive to small-scale surface roughness than to topographic slopes. The radar look direction is from the left. Note that the eastern part of the fresh crater rim at the bottom is distorted due to improper terrain correction.} \label{fig:10}
\end{center}
\end{figure}

\begin{figure}[htbp]
\begin{center}
\includegraphics[scale=0.13]{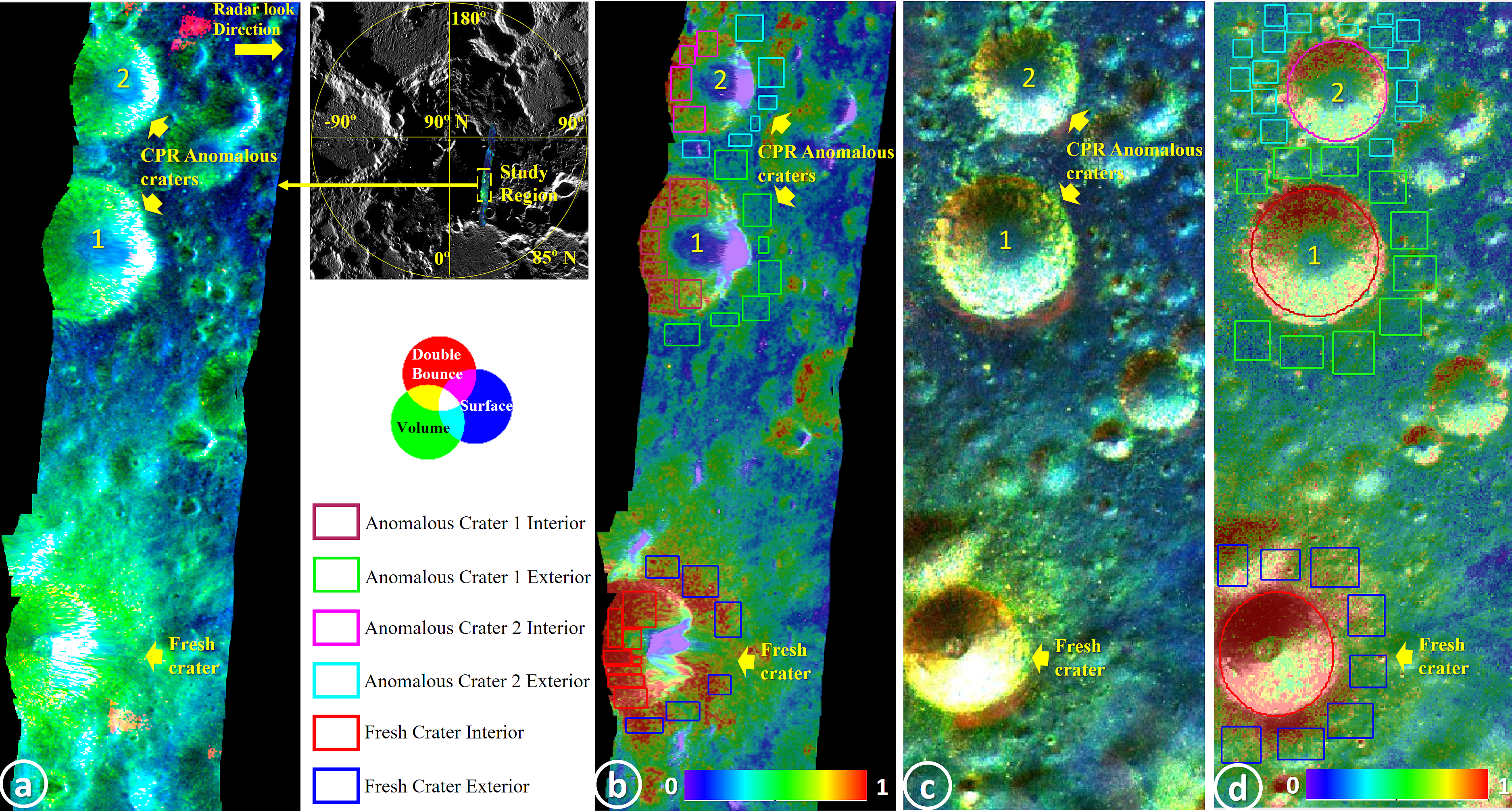}
\caption{Qualitative comparison of secondary craters on the floor of Peary crater from L-band DFSAR and S-band Mini-RF data. (a) L-band DFSAR  Yamaguchi 4-component decomposition image with arrows in yellow denoting the anomalous (labeled ``1”, ``2”) and fresh craters analysed in this work. (b) L-band DFSAR CPR image stretched to a color scale with the interior and exterior regions of interest outlined in different colored boxes as shown in the legend to its left. (c) S-band DFSAR  \textit{m-chi} decomposition image of the same region with arrows in yellow denoting the anomalous and fresh craters. (d) corresponding S-band Mini-RF CPR stretched to a color scale and overlaid on the total backscatter image with the interior and exterior regions of interest. The polygons shown in (b) and (d) are used for the scatter plot shown in figure \ref{fig:12}. North is up and the color scheme for decomposition images is the same as shown in figure \ref{fig:8}. Note that in the case of DFSAR data, we have not sampled the regions affected by artifacts (irregular red patches in the decomposition image (a)) and distorted fresh crater rim region for our statistical analysis.} \label{fig:11}
\end{center}
\end{figure}

L-band data of the Peary crater region was acquired in FP mode at an incidence angle of 26$^{\circ}$ with a resolution (pixel spacing) of $\sim$0.6 m in azimuth and 9.6 m in range. From the calibrated SLC data, we generated the 2$\times$2 scattering matrix followed by the 3$\times$3 [C] matrix (section \ref{sec:4.1}). The averaged [C] matrix was then used with the slant range grid file (section \ref{sec:3.3}) to project the data onto a lunar coordinate system with a pixel spacing of 25 m/pixel. This yielded an $\sim$38-look average for each sampled location in an observation and an approximate 1/N$^{1/2}$ (N=number of looks) uncertainty in the CPR measurements of $\pm$ 0.16 \citep{Campbell_2002}. The process of transforming the unprojected data to a lunar grid also involved correcting the data for terrain height variations that are inherent in the oblique viewing geometry of the SAR instrument. The terrain correction of DFSAR data is performed by utilizing Chandrayaan-2 spacecraft orbit information and (per pixel) topography information available from a global digital elevation model (DEM) at $\sim$ 118 meter spacing available from the LOLA instrument \citep{Mazarico_2011} onboard the LRO mission. Finally, the multilooked, terrain corrected [C] matrix is used to generate the backscatter coefficients, CPR, entropy, average alpha angle, and decomposition images (figure \ref{fig:8}) using equations \ref{eq:7}-\ref{eq:12}, required for our analysis. We analyzed two impact craters on the floor of Peary crater that are classified as “anomalous” in previous studies \citep{Spudis_2010, Spudis_2013, Fa_2018, Virkki_2019} along with a “normal” fresh, young crater to characterize their scattering properties at L-band. Anomalous craters are those having a high CPR in their interior but not exterior to their rims \citep{Spudis_2010, Spudis_2013, Fa_2018}. All three impact craters are within PSRs at the North Pole. Figure \ref{fig:9} shows histograms of CPR, wave entropy and mean alpha angle plotted from the polygons interior and exterior ($\sim$1 crater radii) to the crater rim of these 3 craters as shown in figure \ref{fig:11} b.\\
From Figure \ref{fig:9} (and Table \ref{Table5} in the supporting information), we observe that the two anomalous craters (earlier observed at S-band using Mini-RF data) appear anomalous at L-band also, and the mean CPR and alpha angle values for the young, fresh crater region are relatively higher than those observed for both anomalous craters. The ejecta of young, fresh craters often stand out among other lunar features when observed at radar wavelengths (e.g. \cite{Campbell_2010, Campbell_2012, Raney_2012, Saran_2012}). This is due to the presence of relatively high population of wavelength-scale scatterers found within the ejecta deposits. This roughness signature appears radar-bright in the backscattered power images, associated with high CPR values that exceed unity (Campbell, 2012). Also, rough surfaces are dominated by multiple-bounce backscattering, which randomizes the polarization \citep{Campbell_2002, Campbell_2012}. For this reason, the proximal ejecta region of fresh crater appears distinct in HV polarization (depolarized) image compared to HH and VV polarization images (polarized), shown in figure \ref{fig:10}. Apart from their difference in absolute values (shown in dB scale in figure \ref{fig:10}), the behaviour of anomalous crater regions in all the three polarization channels (HH, HV, VV) appear to be similar.
\FloatBarrier
In contrast with the anomalous craters, the distribution of mean CPR, entropy and alpha angle at the interior and exterior of the fresh crater are almost identical, indicating that the roughness of the two areas is, on average, the same on the scale of centimeters to decimeters. Higher values of mean entropy ($\sim$0.87) and alpha angle ($\sim$41°) clearly indicate the presence of a dominant volume scattering component at regions both inside and outside of the fresh crater. This can be observed qualitatively from figure \ref{fig:11}, which shows the relative contributions of dominant scattering mechansims for each crater region. In addition, we observe a non-zero negligible contribution from the double bounce scattering component ($\sim$4 dB less than the lowest of other two scattering components) and almost no contribution from the helix component of the Yamaguchi 4-component decomposed image of these crater regions. So we neglected the contribution of these two components in this analysis. An interesting observation from the relative contribution of mean surface scattering component from figure \ref{fig:11} a is that the surface scattering distributions of scatterers for all the three crater regions appears similar. As a first approximation, this behaviour may indicate that the depolarization of linear-polarized illuminating signals occurs likely through single scattering by rock edges or other discontinuities, rather than as solely multiple-scattering effects predicted by some analytical models. However, additional terrain parameters such as the rms slope (horizontal scale-dependent) and dielectric properties are required to quantitatively validate this interpretation.
\begin{figure}[htbp!]
\begin{center}
\includegraphics[scale=0.6]{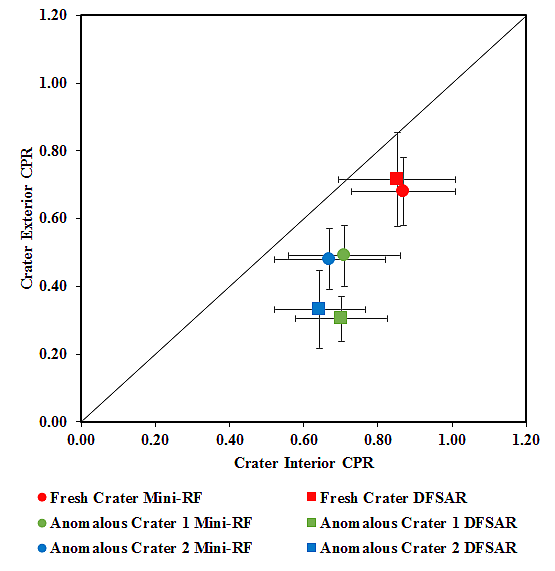}
\caption{Scatter plot of mean CPR values from the interior and exterior regions of two “anomalous” craters and a “normal” fresh crater obtained using S-band Mini-RF and L-band DFSAR quad-pol data. The markers designate data points from the instrument they are collected: DFSAR (squares) and Mini- RF (circles). The colours designate the regions of fresh crater (red), anomalous crater “1” (green), and anomalous crater “2” (blue). Error bars represent one standard deviation of the mean values.} \label{fig:12}
\end{center}
\end{figure}
\FloatBarrier
We then compared our observations with those of S-band Mini-RF data of these crater regions (figure \ref{fig:11} c, d). The Mini-RF level-2 data of this region has a fixed look angle of 49$^{\circ}$ and is reduced to a pixel spacing of 15 m before analysis, giving speckle noise of 1/N$^{1/2}$ $\approx$ 19\% for each pixel. We determined the mean CPR for interior and exterior regions of the three crater regions (from the polygons shown in figure \ref{fig:11} d and values given in Table \ref{Table5} in supporting information) and plotted them against those from DFSAR data for comparison, as shown in figure \ref{fig:12}. This scatter plot reveals an intersting observation – while the mean CPR values for fresh crater interior and ejecta are identical at L- and S-band, they are different for anomalous craters analysed in this study. For the anomalous crater regions, the S-band CPR is higher than that at L-band, implying differences in the material properties and/or rock sizes present in those regions. The difference in the incidence angle between Mini-RF (49$^{\circ}$) and DFSAR data (26$^{\circ}$) does not fully explain this observation. Variations in CPR values for lunar craters are attributed to a combination of radar incidence angle, age, and target materials. For a given crater region, the other two parameters are same, when comparing the CPR response solely as a function of incidence angle. Lunar surface slopes, such as those resulting from crater walls can change the received radar echoes and CPR values since the surface slope affects both the local incidence angle and polarization state of the incident radar wave in the local coordinate frame (e.g. \cite{Fa_2011}). So larger variations in CPR (as a function of incidence angle) are expected within crater interiors, owing to greater variations in the local slopes, compared to their exterior regions. On the contrary, we observed from figure \ref{fig:12} that CPR differences for the exterior regions of the two anomalous craters are relatively higher compared to those of their interiors when observed at the incidence angles of DFSAR (26$^{\circ}$) and Mini-RF (49$^{\circ}$). Therefore, we attribute this result to the change in size distributions of scatterers as a function of wavelength (L- vs. S-band) of observation. If rock size is the dominant effect causing such a discrepancy, radar signals at different wavelengths could be utilized to constrain their sizes and to distinguish surface from buried rocks in the proximal ejecta of craters (e.g. \cite{Ghent_2016}). For the fresh crater region, similar CPR values at both L- and S-band indicate the presence of volumetric scatterers or roughness at cm and larger scales. In the case of the anomalous crater regions, the rocks responsible for S-band enhancement do not cause a similar enhancement in the L-band data, indicating that they are too small to be resolved by the longer-wavelength L-band signals. These observations imply that material evolution of the interior and ejecta of anomalous craters are decoupled, supporting previous interpretations of a large set of lunar craters using S-band Mini-RF data \citep{Fa_2018, Fassett_2018}. Interpreting this decouple phenomena and CPR enhancement seen within the anomalous crater interiors is not the major focus of this work and will be addressed in a future paper.
\subsection{Dual-frequency (L- and S- band) CP data} \label{sec:6.2}
\begin{figure}[htbp]
\begin{center}
\includegraphics[scale=0.5]{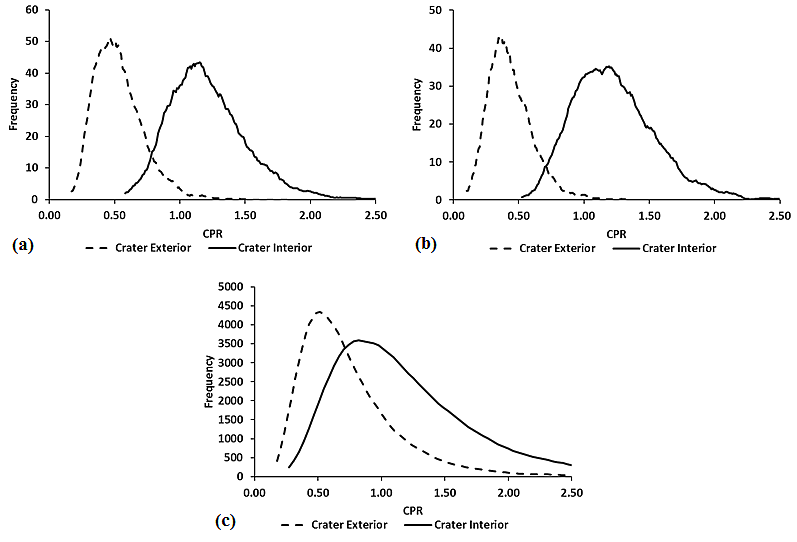}
\caption{CPR histograms for interior and exterior regions of Byrgius C crater derived from (a) S-band DFSAR CP, (b) L-band DFSAR CP, and (c) S-band Mini-RF data.} \label{fig:14}
\end{center}
\end{figure}

We obtained simultaneous dual-frequency (L- and S-band) CP radar data of the crater Byrgius C, which was also classified as “anomalous” in previous studies \citep{Spudis_2013, Fa_2013}. This data was acquired at an incidence angle of 24$^{\circ}$ and has a pixel spacing of $\sim$0.5 m in azimuth and 9.6 m in range. From the calibrated SLC data, we generated the 2$\times$2 covariance matrix for each frequency band and performed multilooking (in the azimuth direction) with N=46 to obtain a slant range pixel spacing of 10 m in range and 24 m in the azimuth direction. We then used this multilooked 2$\times$2 [C] matrix to generate Stokes parameters followed by the CPR and \textit{m-chi} decomposition images for L- and S-band data \citep{Raney_2012}. For this data, the approximate uncertainty in the CPR measurement is $\pm$0.15 for each sampled location.
\begin{figure}[htbp]
\begin{center}
\includegraphics[scale=0.13]{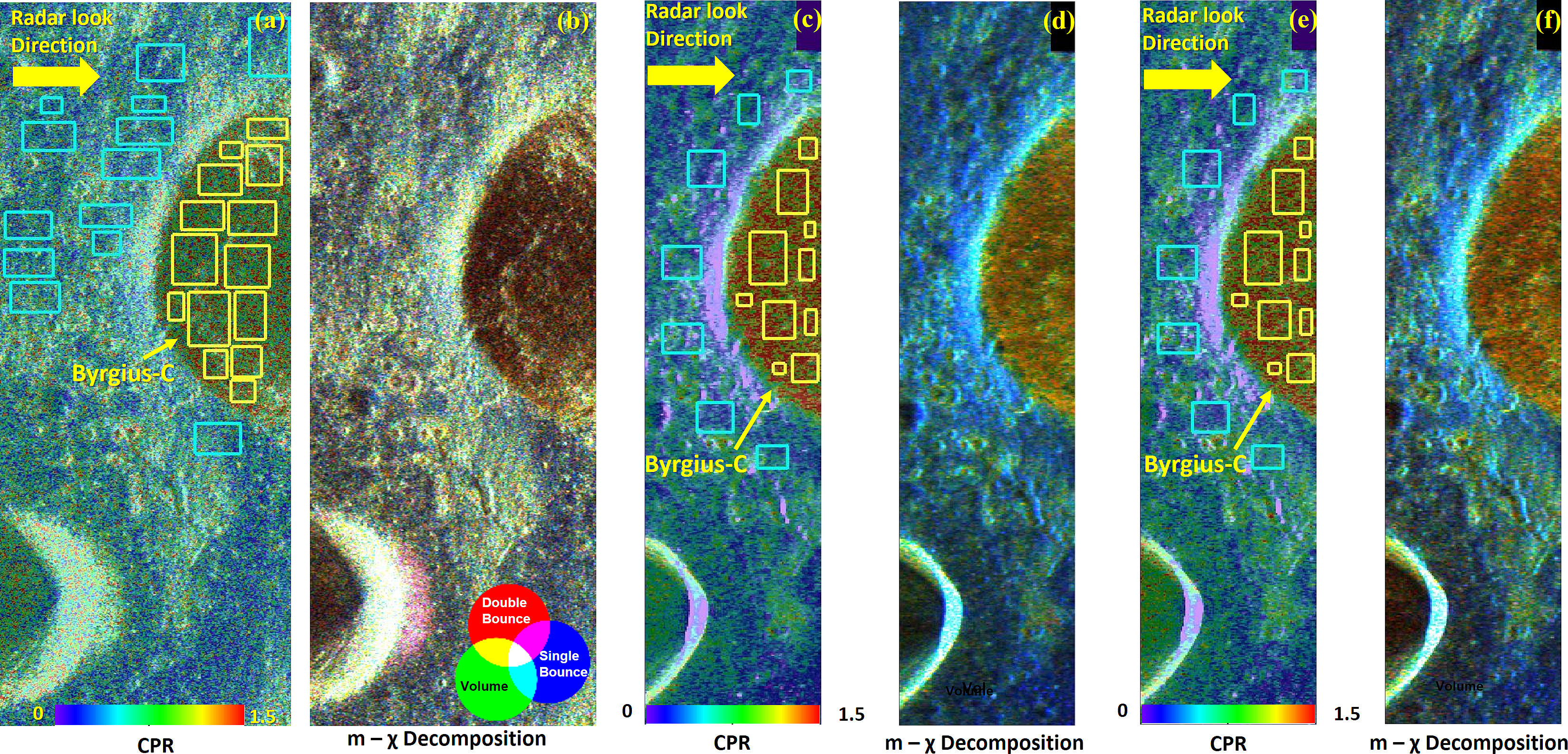}
\caption{Byrgius C crater region observed from the DFSAR and Mini-RF radar instruments. (a) CPR and (b) \textit{m-chi} decomposed image derived from S-band Mini-RF CP data. (c) CPR and (d) \textit{m-chi} decomposed image derived from S-band DFSAR CP data. (e) CPR and (f) \textit{m-chi} decomposed image derived from L-band DFSAR CP data. In the \textit{m-chi} images, red, green, and blue colours represent the double-bounce, random-polarized (volume scattering), and single-bounce component of backscatter respectively. North is up and the radar look direction is indicated in each set of images. All the CPR images are stretched to the same colour scale and overlaid on the total backscattered power images. The polygons shown in (a), (c), and (e) are used to generate the histograms shown in figure \ref{fig:14}. Also note the characteristic red halo \citep{Raney_2012} observed at the far wall (from the radar's perspective) of the crater at the bottom in the Mini-RF \textit{m-chi} decomposed image [(b)] that is missing in the corresponding DFSAR images [(d) and (f)] due to the change in the incidence angle of observation (i.e., 49$^{\circ}$ of Mini-RF vs. 26$^{\circ}$ of the DFSAR)} \label{fig:13}
\end{center}
\end{figure}
\FloatBarrier
Similar to the analysis presented in section \ref{sec:6.1}, we compared our DFSAR observations with those obtained from S-band Mini-RF level-2 data of the Byrgius C crater region (figure \ref{fig:13}). We observe that the mean interior CPRs for Byrgius C exceed 1 and are independent of wavelength (L- vs S-band) and incidence angle (24$^{\circ}$ for DFSAR vs $\sim$50$^{\circ}$ for Mini-RF), contrary to the CPR trends observed from anomalous craters on the floor of Peary crater. The histograms of mean CPRs (Figure \ref{fig:14}) also indicate that the average behavior over the crater interior is CPR $\sim$1 and is limited to less than $\sim$2.5. The mean CPR values (followed by median value in square brackets) determined for the crater interior from DFSAR are 1.17 [1.14] and 1.15 [1.11] for L- and S-bands respectively, compared to 1.16 [1.02] obtained from Mini-RF at S-band. For the ejecta regions, they are 0.45 [0.42] and 0.7 [0.61] from DFSAR (at both L- and S-bands) and Mini-RF respectively. The relatively higher differences between the mean and median values of CPR obtained from Mini-RF data of this region result in a more skewed distribution, as seen in figure \ref{fig:14} c. Even after spatially averaging over a large number of looks to limit speckle effects, there are significant areas of the crater interior with CPR ratios greater than unity observed from both DFSAR and Mini-RF data. These characteristics are consistent with scattering by a combination of the quasi-specular (or small facet) and diffuse mechanisms \citep{Campbell_1993, Campbell_2002}, most possibly due to the presence of a large number of wavelength-sized blocks and/or cracks present at the Byrgius C crater interior. The highest CPR enhancements ($>$2) may be attributable to dihedral mechanism or double-bounce scattering from the randomly oriented dipoles \citep{Campbell_2012}. The \textit{m-chi} decomposition images of the crater interior (Fig. \ref{fig:13} b, d, f) indicate a mixture of volume and double-bounce scattering at work, and support this interpretation.
\FloatBarrier
\subsection{L-band CP data} \label{sec:6.3}
For this analysis, we obtained very high resolution (pixel spacing of $\sim$0.4 m in azimuth and 1.8 m in range) L-band CP data of the Manzinus C crater region at an incidence angle of 30$^{\circ}$. From the calibrated SLC data, we generated the 2$\times$2 covariance matrix and resampled them using the slant range grid file to project the data on to a lunar coordinate system with a pixel spacing of 15 m/pixel. Unlike the previous two comparisons shown in sections \ref{sec:6.1} and \ref{sec:6.2} in which we used level-2 Mini-RF data, we examined Mini-RF data of this region by processing the level-1 data using USGS ISIS3 software as described in \citet{Fa_2018} and \citet{Virkki_2019}. From the original spatial resolution of 7.5 m/pixel of level 1 product, we generated the orthorectified Stokes vector at 15 m/pixel resolution to match with that of the DFSAR L-band data. Finally, the terrain corrected Stokes vector from both datasets is used to calculate the CPR and \textit{m-chi} decomposition images as described in the previous section. Due to the less number of looks, for each pixel in both DFSAR and Mini-RF data used here, the uncertainty in the CPR measurements is on the order of $\pm$0.2.

\begin{figure}[htbp]
\begin{center}
\includegraphics[scale=2]{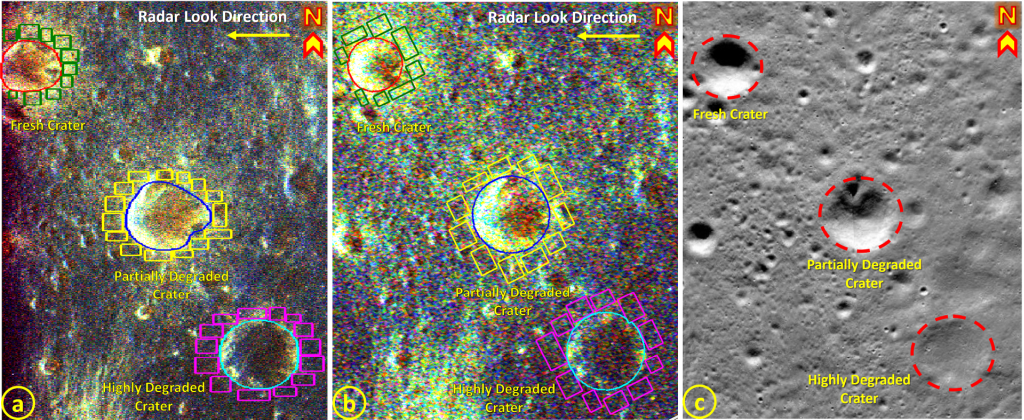}
\caption{\textit{m-chi} decomposition images of three unnamed craters near Manzinus C crater in the lunar south polar region obtained from (a) DFSAR L-band CP data and (b) LRO Mini-RF S-band data; (c) corresponding high resolution LROC NAC image for context (Product ID: M1325822321LE). North is to the top and image centre coordinates are 70.7$^{\circ}$S, 22.5$^{\circ}$E. The color scheme is same as shown in figure \ref{fig:13} and areas of bright yellow within crater regions in the radar images are interpreted to reflect an enhanced abundance of centimeter-to-decimeter rocks. The crater interior (circles and polygons) and exterior (rectangles) regions shown in (a) and (b) are used to generate the histograms shown in figure \ref{fig:16}} \label{fig:15}
\end{center}
\end{figure}

\begin{figure}[htbp]
\begin{center}
\includegraphics[scale=0.6]{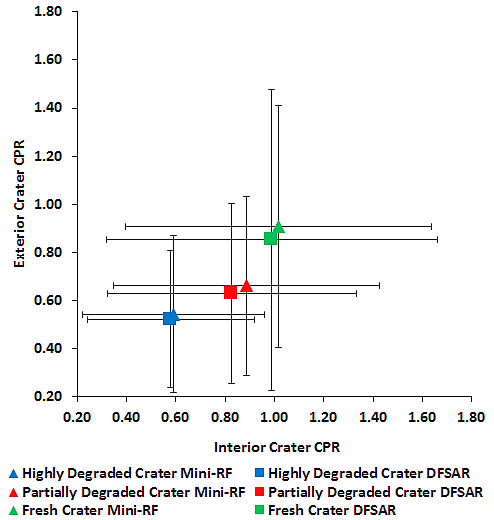}
\caption{Scatter plot of mean CPR values from the interior and exterior regions of three classes of craters (see the legend) near Manzinus C crater, obtained using S-band Mini-RF and L-band DFSAR CP data. The markers designate data points from the instrument they are collected: DFSAR (squares) and Mini-RF (triangles) and the colours designate the regions of highly degraded (blue), partially degraded (red), and fresh crater (green) regions. Error bars represent one standard deviation of the mean values.} \label{fig:16}
\end{center}
\end{figure}

Figure \ref{fig:15} shows the \textit{m-chi} decomposition images obtained from L-band DFSAR and S-band Mini-RF data, compared with a corresponding LROC NAC image. We selected three similar-sized ($\sim$0.8 km) craters and marked them as “fresh”, “partially degraded” and “highly degraded”, based on their roughness properties, evident from a visual inspection of the radar data (figure \ref{fig:15}). We determined the mean CPR for interior and exterior regions of the three craters from the S-band Mini-RF and L-band DFSAR and plotted them as shown in figure \ref{fig:16}. From figures \ref{fig:15} and \ref{fig:16}, we observe that the fresh crater appears distinct at both wavelengths from the other two degraded craters, as the radar signals are sensitive to scattering from surface and buried rocks (size ranges of about 1/10 to 10 wavelengths) and roughness.
A radar-bright ejecta blanket that is excavated by the impact, which is not visible in the corresponding LROC NAC image (figure \ref{fig:15}), surrounds the fresh crater. For the partially degraded crater, which has CPR values intermediate to those of the fresh and highly degraded craters, this radar-bright ejecta halo is not distinct. The CPR values of both the crater interior and proximal ejecta of the highly degraded crater are comparable to the background regolith. CPR measurements obtained at multiple radar wavelengths often provide information about how lunar craters evolve with time, and how rocks at the lunar surface and in the near subsurface breakdown (e.g. \cite{Ghent_2016, Fassett_2018}). A recent model using the S-band CPR and depth/diameter ratio of a set of simple craters suggests that crater degradation state can be used to estimate their age \citep{Fassett_2014}. One of their results suggested that the extent and magnitude of CPR enhancement declines as craters age. For the three crater regions, it is interesting to note that the mean CPR values at their corresponding interior and exterior regions at L- and S-bands are almost similar (figure \ref{fig:16}), given the differences in wavelength and incidence angle. Unlike the results obtained from the comparison of north polar craters as discussed in section \ref{sec:6.1}, the similar CPR values obtained from these three different crater regions suggest the presence of centimeter-to-decimeter blocks. However, the role of the polarimetric basis of observation (FP vs. CP) in determining the CPR needs to be further investigated to fully understand these observed CPR patterns. Nevertheless, if their ages can be calculated by other means (e.g. modelling and crater counting statistics), the distinct CPR patterns between the interior and exterior of these three craters may provide insights into how the rockiness of craters evolves with time.
\FloatBarrier
\section{Conclusions and future scope} \label{sec:7}
DFSAR offers global radar data of the Moon at a range of spatial resolutions to investigate PSRs at the poles and to probe the lunar regolith to reveal their surface and near subsurface physical properties by integrating with other global lunar data sets. The data calibration analysis presented here provides a means to compare on-orbit performance with pre-launch measurements and the results are found to be matched with the pre-launch expected values, meeting their design requirements. By exploiting the full range of information contained in the FP SAR format, our results obtained in this preliminary analysis to decompose lunar radar echoes into physically meaningful components using their polarization properties help to understand the types of surfaces that produce different combinations of polarimetric behaviours. The radar polarization properties of a set of impact craters and their proximal ejecta at different wavelengths provide important information about their physical properties. We observe that craters in both PSRs (Peary secondary craters) and non-PSRs (Byrgius C) that are classified as CPR anomalous in previous S-band radar analyses appear anomalous at L-band also. In the case of Byrgius C crater, we compared our dual-frequency results with the predicted behaviors of scattering by a combination of the small facet and diffuse mechanisms \citep{Campbell_1993, Campbell_2002}, and the highest CPR enhancements ($>$2) are attributed to dihedral mechanism (or double-bounce scattering) from randomly oriented dipoles \citep{Campbell_2012}. Based on our L- and S-band data analysis of crater freshness, further investigation of differences in the CPR evolution of different crater sizes would be a fruitful avenue for future work.

Analysis of FP radar data obtained from DFSAR along with other available radar datasets from the Mini-RF radars (S- and X-band) and Arecibo Observatory (S- and P-band) would enable us to better understand the variability in radar echoes from lunar regolith as a function of wavelength. A major advantage in utilizing the longer wavelength L-band radar is its ability to penetrate the shallow (few-meters at most) regolith , based on the loss tangent, to estimate the influence of regolith composition and (under favourable conditions) the nature of the substrate on the radar echoes. Moreover, the lunar water ice controversy cannot be resolved with existing monostatic data (prior to DFSAR), and new radar measurements are required. The unique dual-wavelength fully polarimetric DFSAR data is better poised to reduce the ambiguities associated with radar detection of water ice at the lunar PSRs, by detecting distinct polarimetric signatures of “dirty ice” regions. Toward this end, constraining the distribution of dielectric properties and surface roughness across the lunar surface, and hence their role in radar surface scattering and subsurface penetration depth, is essential to reducing uncertainties associated with the identification and quantification of potential ice inclusions in the regolith. As linear polarization can provide a quantitative approach to understanding the surface physical properties (i.e., dielectric and textural properties) (e.g. \cite{Campbell_1993, Campbell_2002}), we plan to derive the regolith dielectric constant from DFSAR FP data, followed by utilizing lab-based and theoretical dielectric mixing models (with several ice-regolith mixture combinations) to detect the possible presence of ice inclusions. Furthermore, surface roughness at the PSRs could be estimated by utilizing Mini-SAR, Mini-RF and DFSAR radar data in conjunction, since all these datasets are acquired at different wavelengths and viewing geometry, and therefore offer further validation of centimetre to decimetre near surface roughness.

Finally, with fully polarimetric SAR data of the lunar surface now available for the first time ever from the DFSAR, another significant area of research that could be explored is to investigate whether there is a fundamental difference in the capabilities of different radar architectures (e.g., FP vs. CP) when it comes to measuring the polarimetric properties of a surface, as explored previously in \cite{Carter_2017}. Similar to the analysis presented in \cite{Carter_2017}, we plan to attempt the following comparisons of monostatic radar data by acquiring new DFSAR datasets (at similar viewing geometry) in the near future: (i) DFSAR FP and CP data obtained at the same wavelength, (ii) DFSAR S-band FP and CP data with those of Mini-RF S-band, and (iii) DFSAR S-band FP and CP data with those of Arecibo S-band.

\acknowledgments
The authors acknowledge the contribution of entire DFSAR team responsible for instrument development, characterization, qualification and operations. They would like to thank two anonymous reviewers who provided helpful comments and suggestions that improved this paper.

\bibliography{References}{}

\begin{thebibliography}{}
\expandafter\ifx\csname natexlab\endcsname\relax\def\natexlab#1{#1}\fi
\providecommand{\url}[1]{\href{#1}{#1}}
\providecommand{\dodoi}[1]{doi:~\href{http://doi.org/#1}{\nolinkurl{#1}}}
\providecommand{\doeprint}[1]{\href{http://ascl.net/#1}{\nolinkurl{http://ascl.net/#1}}}
\providecommand{\doarXiv}[1]{\href{https://arxiv.org/abs/#1}{\nolinkurl{https://arxiv.org/abs/#1}}}

\bibitem[{Black {et~al.}(2001)Black, campbell, \& Nicholson}]{Black_2001}
Black, G.~J., campbell, D.~B., \& Nicholson, P.~D. 2001, \icarus, 151, 167,
  \dodoi{10.1006/icar.2001.6616}

\bibitem[{Cahill {et~al.}(2014)Cahill, Thomson, Patterson, Bussey, Neish,
  Lopez, Turner, Aldridge, McAdam, Meyer, Raney, Carter, Spudis, Hiesinger, \&
  Pasckert}]{Cahill_2014}
Cahill, J.~T., Thomson, B., Patterson, G.~W., {et~al.} 2014, \icarus, 243, 173,
  \dodoi{10.1016/j.icarus.2014.07.018}

\bibitem[{Campbell(2002)}]{Campbell_2002}
Campbell, B.~A. 2002, Radar Remote Sensing of Planetary Surfaces (Cambridge
  University Press)

\bibitem[{Campbell(2012)}]{Campbell_2012}
---. 2012, \jgr: Planets, 117, \dodoi{10.1029/2012je004061}

\bibitem[{Campbell {et~al.}(1993)Campbell, Arvidson, \&
  Shepard}]{Campbell_1993}
Campbell, B.~A., Arvidson, R.~E., \& Shepard, M.~K. 1993, \jgr: Planets, 98,
  17099, \dodoi{10.1029/93je01541}

\bibitem[{Campbell \& Shepard(1996)}]{Campbell_1996}
Campbell, B.~A., \& Shepard, M.~K. 1996, \jgr: Planets, 101, 18941,
  \dodoi{10.1029/95je01804}

\bibitem[{Campbell {et~al.}(2010)Campbell, Carter, Campbell, Nolan, Chandler,
  Ghent, Hawke, Anderson, \& Wells}]{Campbell_2010}
Campbell, B.~A., Carter, L.~M., Campbell, D.~B., {et~al.} 2010, \icarus, 208,
  565, \dodoi{10.1016/j.icarus.2010.03.011}

\bibitem[{Campbell {et~al.}(2006)Campbell, Campbell, Carter, Margot, \&
  Stacy}]{Campbell_2006}
Campbell, D.~B., Campbell, B.~A., Carter, L.~M., Margot, J.-L., \& Stacy, N.
  J.~S. 2006, \nat, 443, 835, \dodoi{10.1038/nature05167}

\bibitem[{Campbell {et~al.}(1978)Campbell, Chandler, Ostro, Pettengill, \&
  Shapiro}]{Campbell_1978}
Campbell, D.~B., Chandler, J.~F., Ostro, S.~J., Pettengill, G.~H., \& Shapiro,
  I.~I. 1978, \icarus, 34, 254, \dodoi{10.1016/0019-1035(78)90166-5}

\bibitem[{Carter {et~al.}(2009)Carter, Campbell, Hawke, Campbell, \&
  Nolan}]{Carter_2009}
Carter, L.~M., Campbell, B.~A., Hawke, B.~R., Campbell, D.~B., \& Nolan, M.~C.
  2009, \jgr: Planets, 114, \dodoi{10.1029/2009je003406}

\bibitem[{Carter {et~al.}(2017)Carter, Neish, Nolan, Patterson, Jensen, \&
  Bussey}]{Carter_2017}
Carter, L.~M., Neish, B. A. C. C.~D., Nolan, M.~C., {et~al.} 2017, IEEE
  Transactions on Geoscience and Remote Sensing, 55, 1915,
  \dodoi{10.1109/TGRS.2016.2631144}

\bibitem[{Carter {et~al.}(2012)Carter, Neish, Bussey, Spudis, Patterson,
  Cahill, \& Raney}]{Carter_2012}
Carter, L.~M., Neish, C.~D., Bussey, D. B.~J., {et~al.} 2012, Journal of
  Geophysical Research: Planets, 117, \dodoi{10.1029/2011je003911}

\bibitem[{Chabot {et~al.}(2012)Chabot, Ernst, Denevi, Harmon, Murchie, Blewett,
  Solomon, \& Zhong}]{Chabot_2012}
Chabot, N.~L., Ernst, C.~M., Denevi, B.~W., {et~al.} 2012, \grl, 39,
  \dodoi{10.1029/2012GL051526}

\bibitem[{Chabot {et~al.}(2013)Chabot, Ernst, Harmon, Murchie, Solomon,
  Blewett, \& Denevi}]{Chabot_2013}
Chabot, N.~L., Ernst, C.~M., Harmon, J.~K., {et~al.} 2013, \jgr: Planets, 118,
  26, \dodoi{10.1029/2012JE004172}

\bibitem[{Chabot {et~al.}(2018)Chabot, Shread, \& Harmon}]{Chabot_2018}
Chabot, N.~L., Shread, E.~E., \& Harmon, J.~K. 2018, \jgr: Planets, 123, 666,
  \dodoi{10.1002/2017JE005500}

\bibitem[{Cloude(2009)}]{Cloude_2009}
Cloude, S. 2009, Polarisation: Applications in Remote Sensing (Oxford
  University Press), \dodoi{10.1093/acprof:oso/9780199569731.001.0001}

\bibitem[{Cloude \& Pottier(1996)}]{Cloude_1996}
Cloude, S., \& Pottier, E. 1996, {IEEE} Transactions on Geoscience and Remote
  Sensing, 34, 498, \dodoi{10.1109/36.485127}

\bibitem[{Cloude \& Pottier(1997)}]{Cloude_1997}
---. 1997, {IEEE} Transactions on Geoscience and Remote Sensing, 35, 68,
  \dodoi{10.1109/36.551935}

\bibitem[{Cloude {et~al.}(2012)Cloude, Goodenough, \& Chen}]{Cloude_2012}
Cloude, S.~R., Goodenough, D.~G., \& Chen, H. 2012, {IEEE} Geoscience and
  Remote Sensing Letters, 9, 28, \dodoi{10.1109/lgrs.2011.2158983}

\bibitem[{Cumming \& Wong(2005)}]{Cumming_2005}
Cumming, I.~G., \& Wong, F.~H. 2005, Digital Processing of Synthetic Aperture
  Radar Data: Algorithms and Implementation (Artech House, USA)

\bibitem[{Eke {et~al.}(2014)Eke, Bartram, Lane, Smith, \& Teodoro}]{Eke_2014}
Eke, V.~R., Bartram, S.~A., Lane, D.~A., Smith, D., \& Teodoro, L.~F. 2014,
  \icarus, 241, 66, \dodoi{10.1016/j.icarus.2014.06.021}

\bibitem[{Fa \& Cai(2013)}]{Fa_2013}
Fa, W., \& Cai, Y. 2013, \jgr: Planets, 118, 1582, \dodoi{10.1002/jgre.20110}

\bibitem[{Fa \& Eke(2018)}]{Fa_2018}
Fa, W., \& Eke, V.~R. 2018, \jgr: Planets, 123, 2119,
  \dodoi{10.1029/2018je005668}

\bibitem[{Fa {et~al.}(2011)Fa, Wieczorek, \& Heggy}]{Fa_2011}
Fa, W., Wieczorek, M.~A., \& Heggy, E. 2011, \jgr: Planets, 116,
  \dodoi{10.1029/2010je003649}

\bibitem[{Farr(1993)}]{Farr_1993}
Farr, T.~G. 1993, Radar interactions with geologic surfaces, In: Guide to
  Magellan Image Interpretation (JPL Publication 93-24), 148

\bibitem[{Fassett {et~al.}(2018)Fassett, King, Nypaver, \&
  Thomson}]{Fassett_2018}
Fassett, C.~I., King, I.~R., Nypaver, C.~A., \& Thomson, B.~J. 2018, \jgr:
  Planets, \dodoi{10.1029/2018je005741}

\bibitem[{Fassett \& Thomson(2014)}]{Fassett_2014}
Fassett, C.~I., \& Thomson, B.~J. 2014, \jgr: Planets, 119, 2255,
  \dodoi{10.1002/2014je004698}

\bibitem[{Ghent {et~al.}(2016)Ghent, Carter, Bandfield, Udovicic, \&
  Campbell}]{Ghent_2016}
Ghent, R., Carter, L., Bandfield, J., Udovicic, C.~T., \& Campbell, B. 2016,
  \icarus, 273, 182, \dodoi{10.1016/j.icarus.2015.12.014}

\bibitem[{Ghent {et~al.}(2005)Ghent, Leverington, Campbell, Hawke, \&
  Campbell}]{Ghent_2005}
Ghent, R.~R., Leverington, D.~W., Campbell, B.~A., Hawke, B.~R., \& Campbell,
  D.~B. 2005, \jgr: Planets, 110, \dodoi{10.1029/2004je002366}

\bibitem[{Hagfors(1964)}]{Hagfors_1964}
Hagfors, T. 1964, \jgr: Planets, 69, 3779, \dodoi{10.1029/JZ069i018p03779}

\bibitem[{Hapke(1990)}]{Hapke_1990}
Hapke, B. 1990, \icarus, 88, 407, \dodoi{10.1016/0019-1035(90)90091-m}

\bibitem[{Hapke \& Blewett(1991)}]{Hapke_1991}
Hapke, B., \& Blewett, D. 1991, \nat, 352, 46, \dodoi{10.1038/352046a0}

\bibitem[{Harmon {et~al.}(2001)Harmon, Perillat, \& Slade}]{Harmon_2001}
Harmon, J.~K., Perillat, P.~J., \& Slade, M.~A. 2001, \icarus, 149, 1,
  \dodoi{10.1006/icar.2000.6544}

\bibitem[{Harmon {et~al.}(2011)Harmon, Slade, \& Rice}]{Harmon_2011}
Harmon, J.~K., Slade, M.~A., \& Rice, M.~S. 2011, \icarus, 211, 37,
  \dodoi{10.1016/j.icarus.2010.08.007}

\bibitem[{Harmon {et~al.}(1994)Harmon, Slade, V{\'{e}}lez, Crespo, Dryer, \&
  Johnson}]{Harmon_1994}
Harmon, J.~K., Slade, M.~A., V{\'{e}}lez, R.~A., {et~al.} 1994, \nat, 369, 213,
  \dodoi{10.1038/369213a0}

\bibitem[{Henderson \& Lewis(1998)}]{Henderson_1998}
Henderson, F.~M., \& Lewis, A.~J. 1998, Principles and applications of imaging
  radar. Manual of remote sensing: Third edition, Vol.~2 (Wiley, New York).
\newblock \url{https://www.osti.gov/biblio/293027}

\bibitem[{Lee \& Pottier(2009)}]{LeeJong-Sen2009}
Lee, J.-S., \& Pottier, E. 2009, Polarimetric radar imaging : from basics to
  applications (CRC Press)

\bibitem[{Mazarico {et~al.}(2011)Mazarico, Neumann, Smith, Zuber, \&
  Torrence}]{Mazarico_2011}
Mazarico, E., Neumann, G., Smith, D., Zuber, M., \& Torrence, M. 2011, \icarus,
  211, 1066, \dodoi{10.1016/j.icarus.2010.10.030}

\bibitem[{Mishchenko(1992)}]{Mishchenko_1992}
Mishchenko, M.~I. 1992, Earth Moon and Planets, 58, 127,
  \dodoi{10.1007/bf00054650}

\bibitem[{Neish {et~al.}(2014)Neish, Madden, Carter, Hawke, Giguere, Bray,
  Osinski, \& Cahill}]{Neish_2014}
Neish, C., Madden, J., Carter, L., {et~al.} 2014, \icarus, 239, 105,
  \dodoi{10.1016/j.icarus.2014.05.049}

\bibitem[{Neish {et~al.}(2017)Neish, Hamilton, Hughes, Nawotniak, Garry, Skok,
  Elphic, Schaefer, Carter, Bandfield, Osinski, Lim, \& Heldmann}]{Neish_2017}
Neish, C.~D., Hamilton, C.~W., Hughes, S.~S., {et~al.} 2017, \icarus, 281, 73,
  \dodoi{10.1016/j.icarus.2016.08.008}

\bibitem[{Nozette {et~al.}(1996)Nozette, Lichtenberg, Spudis, Bonner, Ort,
  Malaret, Robinson, \& Shoemaker}]{Nozette_1996}
Nozette, S., Lichtenberg, C.~L., Spudis, P., {et~al.} 1996, Sci, 274, 1495,
  \dodoi{10.1126/science.274.5292.1495}

\bibitem[{Nozette {et~al.}(2010)Nozette, Spudis, Bussey, Jensen, Raney,
  Winters, Lichtenberg, Marinelli, Crusan, Gates, \& Robinson}]{Nozette_2010}
Nozette, S., Spudis, P., Bussey, B., {et~al.} 2010, \ssr, 150, 285,
  \dodoi{10.1007/s11214-009-9607-5}

\bibitem[{Ostro {et~al.}(1992)Ostro, Campbell, Simpson, Hudson, Chandler,
  Rosema, Shapiro, Standish, Winkler, Yeomans, Velez, \&
  Goldstein}]{Ostro_1992}
Ostro, S.~J., Campbell, D.~B., Simpson, R.~A., {et~al.} 1992, \jgr: Planets,
  97, 18227, \dodoi{10.1029/92je01992}

\bibitem[{Putrevu {et~al.}(2016)Putrevu, Das, Vachhani, Trivedi, \&
  Misra}]{Putrevu_2016}
Putrevu, D., Das, A., Vachhani, J., Trivedi, S., \& Misra, T. 2016, Advances in
  Space Research, 57, 627, \dodoi{10.1016/j.asr.2015.10.029}

\bibitem[{Putrevu {et~al.}(2020)Putrevu, Trivedi, Das, Pandey, Mehrotra, Garg,
  Reddy, Gangele, Patel, Sharma, Sijwali, Pandya, Shukla, Seth, Ramanujam, \&
  Kumar}]{Putrevu_2020}
Putrevu, D., Trivedi, S., Das, A., {et~al.} 2020, Current Science, 118, 226,
  \dodoi{10.18520/cs/v118/i2/226-233}

\bibitem[{Raney(2007)}]{Raney_2007}
Raney, R. 2007, {IEEE} Transactions on Geoscience and Remote Sensing, 45, 3397,
  \dodoi{10.1109/tgrs.2007.895883}

\bibitem[{Raney {et~al.}(2012)Raney, Cahill, Patterson, \& Bussey}]{Raney_2012}
Raney, R.~K., Cahill, J. T.~S., Patterson, G.~W., \& Bussey, D. B.~J. 2012,
  \jgr: Planets, 117, n/a, \dodoi{10.1029/2011je003986}

\bibitem[{Raney {et~al.}(2011)Raney, Spudis, Bussey, Crusan, Jensen, Marinelli,
  McKerracher, Neish, Palsetia, Schulze, Sequeira, \& Winters}]{Raney_2011}
Raney, R.~K., Spudis, P.~D., Bussey, B., {et~al.} 2011, Proceedings of the
  IEEE, 99, 808, \dodoi{10.1109/jproc.2010.2084970}

\bibitem[{Rignot(1995)}]{Rignot_1995}
Rignot, E. 1995, \jgr: Planets, 100, 9389, \dodoi{10.1029/95je00485}

\bibitem[{Robinson {et~al.}(2010)Robinson, Brylow, Tschimmel, Humm, Lawrence,
  Thomas, Denevi, Bowman-Cisneros, Zerr, Ravine, Caplinger, Ghaemi, Schaffner,
  Malin, Mahanti, Bartels, Anderson, Tran, Eliason, McEwen, Turtle, Jolliff, \&
  Hiesinger}]{Robinson_2010}
Robinson, M.~S., Brylow, S.~M., Tschimmel, M., {et~al.} 2010, \ssr, 150, 81,
  \dodoi{10.1007/s11214-010-9634-2}

\bibitem[{Saran {et~al.}(2012)Saran, Das, Mohan, \& Chakraborty}]{Saran_2012}
Saran, S., Das, A., Mohan, S., \& Chakraborty, M. 2012, \planss, 71, 18,
  \dodoi{10.1016/j.pss.2012.06.014}

\bibitem[{Simpson \& Tyler(1999)}]{Simpson_1999}
Simpson, R.~A., \& Tyler, G.~L. 1999, \jgr: Planets, 104, 3845,
  \dodoi{10.1029/1998je900038}

\bibitem[{Spudis {et~al.}(2010)Spudis, Bussey, Baloga, Butler, Carl, Carter,
  Chakraborty, Elphic, Gillis-Davis, Goswami, Heggy, Hillyard, Jensen, Kirk,
  LaVallee, McKerracher, Neish, Nozette, Nylund, Palsetia, Patterson, Robinson,
  Raney, Schulze, Sequeira, Skura, Thompson, Thomson, Ustinov, \&
  Winters}]{Spudis_2010}
Spudis, P.~D., Bussey, D. B.~J., Baloga, S.~M., {et~al.} 2010, \grl, 37, n/a,
  \dodoi{10.1029/2009gl042259}

\bibitem[{Spudis {et~al.}(2013)Spudis, Bussey, Baloga, Cahill, Glaze,
  Patterson, Raney, Thompson, Thomson, \& Ustinov}]{Spudis_2013}
---. 2013, \jgr: Planets, 118, 2016, \dodoi{10.1002/jgre.20156}

\bibitem[{Sun {et~al.}(2018)Sun, Li, \& Huang}]{Sun_2018}
Sun, G., Li, Z., \& Huang, L. 2018, IJRS, 40, 3787,
  \dodoi{10.1080/01431161.2018.1552817}

\bibitem[{Thompson {et~al.}(1981)Thompson, Zisk, Shorthill, Schultz, \&
  Cutts}]{Thompson_1981}
Thompson, T., Zisk, S., Shorthill, R., Schultz, P., \& Cutts, J. 1981, \icarus,
  46, 201, \dodoi{10.1016/0019-1035(81)90209-8}

\bibitem[{Thompson {et~al.}(2011)Thompson, Ustinov, \& Heggy}]{Thompson_2011}
Thompson, T.~W., Ustinov, E.~A., \& Heggy, E. 2011, \jgr: Planets, 116,
  \dodoi{10.1029/2009je003368}

\bibitem[{Thomson {et~al.}(2012)Thomson, Bussey, Neish, Cahill, Heggy, Kirk,
  Patterson, Raney, Spudis, Thompson, \& Ustinov}]{Thomson_2012}
Thomson, B.~J., Bussey, D. B.~J., Neish, C.~D., {et~al.} 2012, \grl, 39, n/a,
  \dodoi{10.1029/2012gl052119}

\bibitem[{Virkki \& Bhiravarasu(2019)}]{Virkki_2019}
Virkki, A.~K., \& Bhiravarasu, S.~S. 2019, \jgr: Planets, 124, 3025,
  \dodoi{10.1029/2019je006006}

\bibitem[{Yamaguchi {et~al.}(2005)Yamaguchi, Moriyama, Ishido, \&
  Yamada}]{Yamaguchi_2005}
Yamaguchi, Y., Moriyama, T., Ishido, M., \& Yamada, H. 2005, ITGRS, 43, 1699,
  \dodoi{10.1109/tgrs.2005.852084}

\bibitem[{Zyl {et~al.}(1987)Zyl, Zebker, \& Elachi}]{van_Zyl_1987}
Zyl, J. J.~V., Zebker, H.~A., \& Elachi, C. 1987, Radio Science, 22, 529,
  \dodoi{10.1029/rs022i004p00529}

\end{thebibliography}
\bibliographystyle{aasjournal}


\appendix
\section{Supporting information}
\begin{figure}[htbp]
\begin{center}
\includegraphics[scale=1]{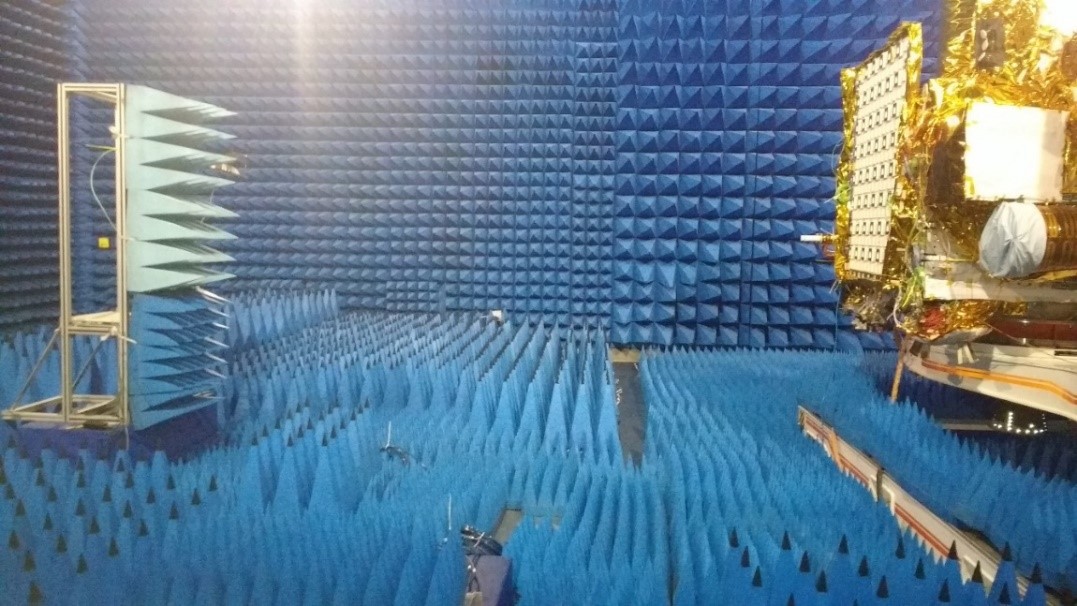}
\caption{Axial ratio measurements in anechoic chamber in fully integrated condition of payload over the spacecraft (shown on the right of the figure). Towards the left is linearly polarized horn antenna used to measure the signal intensity with different orientation angles.} \label{fig:17}
\end{center}
\end{figure}
\begin{table}[htbp] 
\caption{CPR, Entropy and mean Alpha angle values derived from DFSAR L-band FP data of secondary craters on the floor of Peary crater. For comparison, mean CPR values derived from corresponding LRO Mini-RF S-band data are listed here. These values represent the average of profiles analysed from the regions interior and exterior ($\sim$1.5 radii) to the crater rims indicated in Figure \ref{fig:12}, with one sigma standard deviation given as error for CPR.} 
\label{Table5}
\centering
 \begin{tabular}{|c c c c c c|} 
  \hline \hline
 \multicolumn{2}{c}{\textbf{Region}} & \textbf{CPR} & \textbf{CPR} & \textbf{Entropy} & \textbf{Mean Alpha}\\ 
 & &  (DFSAR) & (Mini-RF) &  & (degree) \\ \hline
 Fresh Crater &	exterior &	0.72 $\pm$ 0.14	& 0.68 $\pm$ 0.10 & 0.86 & 38.8 \\ \cline{2-6}
              &	interior &	0.82 $\pm$ 0.16	& 0.87 $\pm$ 0.14 & 0.89 & 42.5 \\ \hline
 Anomalous    &	exterior &	0.30 $\pm$ 0.08	& 0.49 $\pm$ 0.09 & 0.62 & 23.1 \\ \cline{2-6}
 crater 1     &	interior &	0.70 $\pm$ 0.13	& 0.71 $\pm$ 0.15 & 0.85 & 38.7 \\ \hline
 Anomalous    &	exterior &	0.33 $\pm$ 0.09	& 0.48 $\pm$ 0.09 & 0.63 & 24 \\ \cline{2-6}
 crater 2     &	interior &	0.64 $\pm$ 0.13	& 0.67 $\pm$ 0.15 & 0.83 & 36.9 \\ \hline
    \end{tabular}
\end{table}
\end{document}